\documentclass[11pt,a4 paper]{article}
\newtheorem{theo}{Theorem}[section]
\newtheorem{cor}{Corollary}
\newtheorem{rem}{Remark}[section]
\newtheorem{defi}{Definition}[section]
\newtheorem{lemma}{Lemma}[section]

\newtheorem{prop}{Proposition}[section]
\newtheorem{ex}{Example}[section]

\newtheorem{cons}{Construction}

\newcommand{\proof}{\noindent{\em Proof.}\quad}

\usepackage{multirow,bm}
\usepackage{hhline}
\usepackage{array}
\usepackage{makecell}
\newcommand{\F}{\mathbb{F}}

\usepackage{float}
\usepackage{tocloft}
\usepackage{amsfonts,amsmath,amssymb}
\usepackage{multirow, verbatim}
\usepackage{shadow,enumerate}
\usepackage{makeidx}  
\usepackage{graphicx}
\usepackage{color}

\usepackage{mathrsfs}

\newcommand{\FB}{{\mathbb F}_{2}}
\def\whitebox{{\hbox{\hskip 1pt
        \vrule height 6pt depth 1.5pt
        \lower 1.5pt\vbox to 7.5pt{\hrule width
                  3.2pt\vfill\hrule width 3.2pt}%
        \vrule height 6pt depth 1.5pt
        \hskip 1pt } }}
\def\qed{\ifhmode\allowbreak\else\nobreak\fi\hfill\quad\nobreak\whitebox\medbreak}

\begin{document}
\title{A general framework for  secondary constructions \\ of bent and plateaued functions}

\author{
S. Hod\v zi\'c \footnote {University of Primorska, FAMNIT, Koper,
Slovenia, e-mail: samir.hodzic@famnit.upr.si}\and E.~Pasalic\footnote{
University of Primorska, FAMNIT \& IAM, Koper, Slovenia, e-mail: enes.pasalic6@gmail.com}
\and Y. Wei \footnote{
 Guilin University of Electronic Technology, Guilin,
P.R. China,
 e-mail: walker$_{-}$wyz@guet.edu.}
}

\date{}
\maketitle

\begin{abstract}
In this work, we employ the concept of {\em composite representation} of Boolean functions, which represents  an arbitrary Boolean function  as a composition of one  Boolean function and one vectorial function, for the purpose of specifying new secondary constructions of bent/plateaued functions.
This representation gives a better understanding of the existing  secondary constructions  and it also allows us to provide a general construction framework of these objects. This framework   essentially gives rise to  an {\em infinite number} of possibilities  to specify  such secondary construction methods (with some induced sufficient conditions imposed on initial functions) and in particular we solve several open problems in this context.
We provide several  explicit methods  for specifying new   classes of bent/plateaued functions and demonstrate   through examples that  the imposed initial conditions can be easily satisfied.  Our approach is especially efficient when defining new bent/plateaued functions  on  larger variable spaces than initial functions. For instance, it is shown that the indirect sum methods and Rothaus' construction are just special cases of this general framework and some explicit extensions of these methods are given.
In particular, similarly to the basic indirect sum method of Carlet, we show that it is possible to derive (many) secondary constructions of bent functions without any additional condition on initial functions apart from the requirement that these are bent functions.
In another direction, a few construction methods that generalize   the secondary constructions which do not extend the variable space of the employed initial functions   are also proposed.
%

\end{abstract}

\section{Introduction}


Introduced by O. S. Rothaus in 1976, \emph{bent} (or maximally non-linear) functions became one of the most interesting and important combinatorial objects, due to their wide range of applications (for instance coding theory, difference set theory, cryptography).
During the last four decades bent functions have been intensively studied, which resulted in several general (primary) classes: Maiorana-McFarland class ($\mathcal{MM}$) \cite{MMclass}, Partial spread class ($\mathcal{PS}$)  of Dillon \cite{Dillon}, and Dobbertin's $\mathcal{H}$ class \cite{Do95}.
A somewhat related class of functions, characterized by the property that their Walsh spectrum is three-valued (more precisely $0,\pm 2^r$ for a positive integer $r$), is the class of \emph{plateaued} functions which has been introduced in \cite{Zheng} and  later studied in \cite{CarletAPN,CarletQ,Zheng2,Wang,Jong,SihemPlate}. The notion of plateaued functions as defined here does not include bent and linear functions (which only have  two different values in their Walsh spectrum), though sometimes in the literature these families are also included.

 Primary construction methods (referring mainly to bent functions), also referred to as direct construction methods,   employ a collection of suitable algebraic structures on $n/2$-dimensional subspaces  rather than using known   bent functions as their building blocks.
On the other hand, secondary constructions use some initial functions (mainly bent or plateaued) that satisfy certain conditions for the purpose of constructing new bent functions.
In the literature, the bent functions  obtained by secondary constructions   are commonly defined on larger variable spaces, though  alternatively they may be defined on the same variable space as the initial functions. Nevertheless, only  a few  constructions of the latter type are known, for instance see \cite[Section 6.4.2]{CarletBoolean} and \cite[Chapter 6]{SihemBook}). When a classification  of bent functions is considered (which seems to be quite illusive today), the primary construction methods are more important than the secondary ones.   This is because, in general, it is not clear whether these secondary constructions are simply embedded in some of the known primary classes.  However, there is some evidence that suitably chosen initial functions may give rise to bent functions that are not included in the completed versions of primary classes (viewed as a global affine equivalent class of the initial primary class)  such as the $\mathcal{D}_0$ class of Carlet \cite{CarletTNC} and  certain bent functions that origin  from the method of Rothaus \cite{RothEnes2016}.
The importance of these  secondary constructions has also been acknowledged in many recent works, see for instance \cite{CarletRes,CarletGPS,CarletPiece,CarletQ,CarletN,CarletMess,CarletRSF,CarletRESB,Dobbertin,Wilfried,Ayca,Feng,Feng2,Sihem,Sihem2,Xiang, Feng4, RothEnes2016}.

In this work we
 introduce an alternative method of specifying (infinitely) many secondary constructions of bent/plateaued functions by employing a composite representation of Boolean functions.
More precisely, whereas an arbitrary Boolean function $\mathfrak{f}(x)=\mathfrak{f}(x_1,\ldots,x_n)$ is commonly represented by its ANF (see (\ref{ANF})), an alternative way is to define $\mathfrak{f}$
as a composition of one Boolean function $f:\mathbb{F}^k_2 \rightarrow \mathbb{F}_2$ and one vectorial function $H:\mathbb{F}^n_2 \rightarrow \mathbb{F}^k_2$ so that
\begin{eqnarray}\label{F}
\mathfrak{f}(x)=f(H(x))=f(h_1(x),\ldots,h_k(x)),
\end{eqnarray}
where $h_i:\mathbb{F}^n_2 \rightarrow \mathbb{F}_2$,  $i\in[1,k]$,  are called the \emph{coordinate} functions of $H(x)=(h_1(x),\ldots,h_k(x)).$ Throughout the article, the function $f:\mathbb{F}^k_2 \rightarrow \mathbb{F}_2$ is said to be a \emph{form} of  $\mathfrak{f}$\footnote{For convenience and brevity the function $f$ is simply called ''form'' though an alternative way may be to call it "outer function" which is a standard terminology used in coding theory for instance.}.  
 The core idea of this representation (although it is \emph{not unique}, cf. Section \ref{sec:formulas}), is to replace the linear coordinates $x_1,\ldots,x_k$ of $f$ by  new "coordinates" $h_1,\ldots,h_k$ which are not necessarily linear.

Based on this representation, we provide a general framework for specifying a  variety of  secondary construction methods  of bent and plateaued functions.
 We actually generalize a vast majority of secondary constructions (mainly given in \cite{CarletRes,CarletPiece,CarletQ,CarletRSF,Xiang,CarletRESB,CarletGPS,Feng,Feng2,Sihem,Sihem2}), including indirect sums \cite{SecEnf,CarletRESB} and the  construction of Rothaus \cite{Rot},   and thereby solve some open problems that regard finding new efficient secondary constructions (Open Problem 15  in \cite[Section 4.5]{CarletOP}).
It is important to notice that  the composite representation appears to be highly efficient in this context   mainly due to the fact that the integers $k$ and $n$ are not related to  each other which gives a lot of freedom to select coordinate functions $h_1,\ldots,h_k$ appropriately.
The efficiency and flexibility of our method is essentially based on an inherent two-step design process. More precisely, one firstly  constructs a suitable form  $f$ (mainly plateaued or bent) which  induces certain bent (plateaued) conditions. Then,  the coordinate functions $h_i$ that satisfy these conditions are specified. This approach also gives a much better insight and understanding of various known secondary constructions along with the possibility to easily identify suitable modifications of these methods.

The main result  of this article is a general framework for defining new secondary constructions of bent (plateaued) functions, regardless of whether the variable space of generated functions is increased or not.  In the former case, referring to methods that generate new bent (plateaued) functions on  larger variable spaces than initial functions, our main contributions are summarized as follows:
 \begin{itemize}
 \item Generic construction methods of new indirect sums that use a large number of initial bent/plateaued functions are provided in Section \ref{sec:sepvar}. In particular, in Section \ref{subsec:genericRot} we show that the Rothaus construction \cite{Rot} can be efficiently
 generalized in many different ways, e.g. Theorem \ref{theo:rotcompl}, Theorem \ref{afterrot} and Corollary \ref{cor:bent} give explicit design methods.

\item 
    An Open Problem 13 on generalizing the indirect sum without initial conditions \cite[Section 4.5]{CarletOP}, is solved in Section \ref{sec:genindsum} by providing three explicit construction methods, see Theorem~\ref{newindsum} - \ref{newindsuM}.

%


\item Another interesting issue, Open Problem 7 in \cite[Section 4.6]{CarletOP}, regarding the construction of  semi-bent functions from the known ones  is solved in Sections \ref{sec:linear} and \ref{sec:dual2}.
\item Some  efficient construction methods  of plateaued Boolean functions  are given in Section~\ref{sec:plateWS} (also in Sections \ref{sec:sepvar} and \ref{sec:dual2}),   which was posed as an open problem  in \cite{CarletAPN}. These methods specify plateaued functions in their spectral domain which substantially differ from the existing approaches that mainly employ the algebraic normal form domain.


\end{itemize}


In another direction, we also use the  compositional form to address the problem of finding  new secondary constructions of bent/plateaued functions on the same variable space, see Section \ref{sec:bent}.
 We show that  a great variety of constructions is again possible and apart from a general design framework we  give some explicit construction methods (cf. Section \ref{sec:dual2})
  that use an indicator set as the form $f$. As already remarked, this type of constructions  
 seems to be  intrinsically harder  (only a few methods are known \cite{CarletRes,Sihem,Sihem2}) than the methods that extend the variable space.

It is beyond the scope of this work to examine whether the resulting bent  functions (in Section \ref{sec:sepvar} and Section \ref{sec:bent}) are in general contained in the completed versions of the known primary classes. This question is intrinsically hard and cannot be addressed properly since we provide generic design methods and depending on the choice of initial functions the resulting bent function may or may be not included in the known primary classes. In this context we provide an evidence that within a certain class of bent functions (corresponding to one extension of Rothaus method) there are examples of bent functions outside the $\mathcal{MM}$ class (cf. Example \ref{ex:4.7proof}).
Notice that for $n \leq 6$ all bent functions are contained in the completed $\mathcal{MM}$ class \cite{Dillon2}, whereas for $n\geq 8$ the criterion (based on the second order derivatives) for deciding whether a given bent function is outside the completed $\mathcal{MM}$ class cannot be efficiently conducted by a computer due to high computational complexity.  Nevertheless, this question has a great importance since a suitable choice of initial functions may generate bent functions outside the completed versions of known primary classes, see  for instance the lengthy analysis in \cite{RothEnes2016,OutsideMM}.

On the other hand, we were able to show   that at least some of these generalizations are affine inequivalent to the original methods, whereas for  other generic  methods given in this article we provide some informal arguments that our methods cannot be viewed as a simple extension of the known methods. These informal arguments are based on   considering suitable restrictions of our methods (thus fixing a certain subset of input variables) and comparing these restrictions to the indirect sum or Rothaus method. In general, our methods offer much more variety since they involve more initial functions. Moreover, the analysis of these restrictions to certain (affine) subspaces reveals some substantial differences compared to the known methods. In particular, unlike the method of Rothaus the restrictions of our methods  are generally not bent. Furthermore, for some of our  generalizations  these restrictions stem from both the direct and indirect sum method which makes them interesting in many contexts.

\section{Preliminaries}\label{sec:pre}

The vector space $\mathbb{F}_2^n$ is the space of all $n$-tuples $x=(x_1,\ldots,x_n)$, where $x_i \in \mathbb{F}_2$.
For $x=(x_1,\ldots,x_n)$ and $y=(y_1,\ldots,y_n)$  in $\mathbb{F}^n_2$, the usual scalar (or dot) product over $\mathbb{F}_2$ is defined as $x\cdot y=x_1 y_1\oplus\cdots\oplus x_n y_n.$ The Hamming weight of  $x=(x_1,\ldots,x_n)\in \mathbb{F}^n_2$ is denoted and computed as  $wt(x)=\sum^n_{i=1} x_i.$ By "$\sum$" we denote the integer sum (without modulo evaluation), whereas "$\bigoplus$" denotes the sum evaluated modulo two.

The set of all Boolean functions in $n$ variables, which is the set of mappings from $\mathbb{F}_2^n$ to $\mathbb{F}_2$, is denoted by $\mathcal{B}_n$.  Especially, the set of affine functions in $n$ variables is given by $\mathcal{A}_n=\{a\cdot x\oplus b\;|\;a\in\mathbb{F}_2^n,\; b\in\{0,1\}\},$ and similarly  $\mathcal{L}_n=\{a\cdot x:a\in\mathbb{F}_2^n\}\subset \mathcal{A}_n$ 
 denotes the set of linear functions. It is well-known that any $f:\mathbb{F}^n_2 \rightarrow \mathbb{F}_2$ can be uniquely represented by its associated algebraic normal form (ANF) as follows:
\begin{eqnarray}\label{ANF}
f(x_1,\ldots,x_n)={\bigoplus_{u\in \mathbb{F}^n_2}{\lambda_u}}{(\prod_{i=1}^n{x_i}^{u_i})},
\end{eqnarray}
where $x_i, \lambda_u \in \mathbb{F}_2$ and $u=(u_1, \ldots,u_n)\in \mathbb{F}^n_2$. The support of an arbitrary function $f\in \mathcal{B}_n$ is defined as $supp(f)=\{x\in \mathbb{F}^n_2:f(x)=1\}.$ 

For an arbitrary function $f\in \mathcal{B}_n$, the set of its values on $\mathbb{F}^n_2$ (\emph{the truth table}) is defined as $T_f=(f(0,\ldots,0,0),f(0,\ldots,0,1),f(0,\ldots,1,0),\ldots,f(1,\ldots,1,1))$. The corresponding $(\pm 1)$-{\em sequence of $f$} is defined as
$\chi_f=((-1)^{f(0,\ldots,0,0)},(-1)^{f(0,\ldots,0,1)},(-1)^{f(0,\ldots,1,0)}\ldots,$ $(-1)^{f(1,\ldots,1,1)})$. The {\em Hamming distance} $d_H$ between two arbitrary Boolean functions, say $f,g\in \mathcal{B}_n,$ we define by $d_H(f,g)=\{x\in \mathbb{F}^n_2:f(x)\neq g(x)\}=2^{n-1}-\frac{1}{2}\chi_f\cdot \chi_g$, where $\chi_f\cdot \chi_g=\sum_{x\in \mathbb{F}^n_2}(-1)^{f(x)\oplus g(x)}$.

 For any two sets $A=\{\alpha_1,\ldots,\alpha_{r}\}$ and $B=\{\beta_1,\ldots,\beta_{r}\}$, let $A\wr B=\{(\alpha_i,\beta_i): i=1,\ldots,r\}$. For arbitrary sized sets of binary vectors, the Kronecker product of $A$ and $B$ is $A\times B=\{(\alpha,\beta):\alpha\in A,\; \beta\in B\}$. By $\textbf{0}_k=(0,\ldots,0)$ we denote the all-zero vector in $\mathbb{F}^k_2.$

The \emph{Walsh-Hadamard transform} (WHT) of $f\in\mathcal{B}_n$, and its inverse WHT, at any point $\omega\in\mathbb{F}^n_2$ are defined, respectively,  by
\begin{eqnarray}\label{WHT}
W_{f}(\omega)=\sum_{x\in \mathbb{F}_2^n}(-1)^{f(x)\oplus \omega\cdot x},\;\;\;\;(-1)^{f(x)}=2^{-n}\sum_{\omega\in \mathbb{F}_2^n}W_f(\omega)(-1)^{\omega\cdot x}.
\end{eqnarray}
A \emph{vectorial Boolean function}, say $H:\mathbb{F}^n_2\rightarrow\mathbb{F}^k_2$, can be represented uniquely as $H(x)=(h_1(x),\ldots,h_k(x))$, where $h_i:\mathbb{F}^n_2 \rightarrow \mathbb{F}_2$. For any fixed non-zero vector $\omega\in \mathbb{F}^k_2$ and any $u\in \mathbb{F}^n_2,$ the notation $W_{\omega\cdot H}(u)$ will denote the WHT  of the  {\em component} function  $\omega\cdot H=\omega_1h_1 \oplus \cdots \oplus \omega_k h_k$ of the function $H$ at point $u$.

For any subset $U\subseteq \mathbb{F}^n_2,$  $\phi_U$ will denote a Boolean function for which $\phi_U(x)=1$ if and only if $x\in U.$ Also, by $U^{\bot}$ we denote the set  $U^{\bot}=\{y\in \mathbb{F}^n_2:x\cdot y=0,\;\forall x\in U\}.$ The cardinality of any set $U$ is  denoted by  $\#U.$


\subsection{Bent and plateaued functions and their duals}

Throughout this article we use the following definitions related to bent and plateaued functions:
\begin{itemize}
\item A function $f\in\mathcal{B}_n,$ for even $n$, is called {\em bent} if $W_f(u)=2^{\frac{n}{2}}(-1)^{f^*(u)}$
for a Boolean function $f^*\in \mathcal{B}_n$ which is also a bent function, called the {\it dual} of $f$.


\item Two functions $f$ and $g$ on $\FB^n$ are said to be at {\em bent distance} if $d_H(f,g)= 2^{n-1}\pm 2^{n/2-1}$. Similarly, for a subset $B\subset \mathcal{B}_n$, a function $f$ is said to be at bent distance to $B$ if for all $g\in B$ it holds that $d_H(f,g)=2^{n-1}\pm 2^{n/2-1}$.

\item A function $f\in \mathcal{B}_n$ is called {\em $s$-plateaued} if its Walsh spectrum only takes three values $0$ and $\pm 2^{\frac{n+s}{2}}$ ($\leq 2^n$), where $s\geq 1$ if $n$ is odd and $s\geq 2$ if $n$ is even ($s$ and $n$ always have the same parity). The Walsh distribution of $s$-plateaued functions (cf. \cite[Proposition 4]{Decom}) is given by
\begin{eqnarray*}
\begin{tabular}{|c|c|}
  \hline
  $W_f(u)$ & Number of $u\in \mathbb{F}^n_2$ \\ \hline
  $0$ & $2^n-2^{n-s}$ \\ \hline
  $2^{\frac{n+s}{2}}$ & $2^{n-s-1}+(-1)^{f(0)}2^{\frac{n-s}{2}-1}$ \\ \hline
  $-2^{\frac{n+s}{2}}$ & $2^{n-s-1}-(-1)^{f(0)}2^{\frac{n-s}{2}-1}$ \\
  \hline
\end{tabular}
\end{eqnarray*}
In particular, a class of $1$-plateaued functions for $n$ odd, or  $2$-plateaued for $n$ even, corresponds to so-called {\em semi-bent} functions.

\item The {\em Walsh support} of  $f\in \mathcal{B}_n$ is defined as $S_f=\{\omega\in \mathbb{F}^n_2\; :\; W_f(\omega)\neq0\}.$

 \item For an arbitrary  $s$-plateaued function $f\in \mathcal{B}_n$ with $W_f(u)\in \{0,\pm 2^{\frac{n+s}{2}}\}$ the value $2^{\frac{n+s}{2}}$ is called the \emph{amplitude} of $f$. We define its dual function $f^*$ on the set $S_f$ of cardinality $2^{n-s}$ by $W_f(\omega)=2^{\frac{n+s}{2}}(-1)^{f^*(\omega)},$ for $\omega\in S_f$\footnote{In some other articles, the dual of plateaued functions is defined by including $W_f(\omega)=0$ on $\mathbb{F}^n_2\backslash S_f$, i.e., $f^*(\omega)=0$ if $W_f(\omega)=0$, and $f^*(\omega)=1$ if $W_f(\omega)\neq 0$ ($\omega\in \mathbb{F}^n_2$). However, this definition of dual $f^*$ does not distinguish positive and negative values of non-zero coefficients $W_f(\omega)$, neither it uniquely determines the corresponding plateaued function $f$.}.
     %
     %
To specify the dual function as $f^*:\FB^{n-s} \rightarrow \FB$ we use the concept of {\em lexicographic ordering}. That is, a subset  $E=\{e_0,\ldots,e_{2^{n-s}-1}\}\subset \mathbb{F}^{n}_2$ is ordered lexicographically if $|e_i| < |e_{i+1}|$ for any $i \in [0,2^{n-s}-2]$, where $|e_i|$ denotes the integer representation of $e_i \in \mathbb{F}^n_2$. More precisely, for $e_i=(e_{i,0}, \ldots, e_{i,n-1})$ we have $|e_i|= \sum_{j=0}^{n-1}e_{i,n-1-j} 2^j$, thus having the most significant bit of $e_i$ on the left-hand side. Since $S_f$ is not ordered in general, {\em we will always represent it} as $S_f=v \oplus E$, where $E$ is lexicographically ordered for some fixed  $v \in S_f$ and $e_0=\textbf{0}_{n}$. For instance, if $S_f=\{(0,1,0),(0,1,1),(1,0,0), (1,0,1)\}$, by fixing $v=(0,1,1)\in S_f$, then $E=\{e_0,e_1,e_2,e_3\}=\{(0,0,0), (0,0,1),(1,1,0),(1,1,1)\}$ is ordered lexicographically and consequently $S_f$ is "ordered" as $S_f=\{\omega_0,\omega_1,\omega_2,\omega_3\}=\{(0,1,1),(0,1,0),(1,0,1),(1,0,0)\}$. 

 This way we can make a direct correspondence between  $\FB^{n-s}$ and $S_f$ through $E$ so that for $\FB^{n-s}=\{x_0, x_1, \ldots, x_{2^{n-s}-1}\}$, where $\FB^{n-s}$ is lexicographically ordered, we have
\begin{eqnarray}\label{DPL}
 f^*(\omega_j)\leftrightsquigarrow f^*(e_j)\leftrightsquigarrow f^*(x_j),\;\;\; x_j \in \FB^{n-s}, \; e_j \in E,\;j\in[0,2^{n-s}-1],
 \end{eqnarray}
i.e., we set that $S_f=\{\omega_0,\ldots,\omega_{2^{n-s}-1}\}$ is ordered so that $\omega_i=v\oplus e_i$, and $E=\{e_0,\ldots,e_{2^{n-s}-1}\}$ is ordered lexicographically.  In the above example we have for instance that
\begin{table}[h!]
\scriptsize
\centering
\begin{tabular}{|c|c|c|}
  \hline
  \makecell{$S_f=v\oplus E$ given with respect to\\lexicographically ordered\\ $E$ for $v=(0,1,1)$} & \makecell{The lexicographically\\ ordered set $E$} & \makecell{Values of $f^*(\omega_i)=f^*(e_i)=f^*(x_i)$,\\
  where $x_i\in \mathbb{F}^{2}_2$\\ $(\mathbb{F}^{2}_2$ is ordered lexicographically)} \\ \hline
  $\omega_0=(0,1,1)$ & $(0,0,0)=e_0$ & $f^*(\omega_0)=f^*(x_0)=f^*(0,0)$ \\
  $\omega_1=(0,1,0)$ & $(0,0,1)=e_1$ & $f^*(\omega_1)=f^*(x_1)=f^*(0,1)$ \\
  $\omega_2=(1,0,1)$ & $(1,1,0)=e_2$ & $f^*(\omega_2)=f^*(x_2)=f^*(1,0)$ \\
  $\omega_3=(1,0,0)$ & $(1,1,1)=e_3$ & $f^*(\omega_3)=f^*(x_3)=f^*(1,1)$ \\
  \hline
\end{tabular}
\caption{Values of $f^*$ with respect to $S_f$.}
\label{tab1}
\end{table}
\begin{rem}
Formally, one can specify  $P$ to be a mapping from $S_f=v\oplus E$ to $\mathbb{F}^{n-s}_2$ ($E=\{e_0,\ldots,e_{2^{n-s}-1}\}$ is ordered lexicographically, $\omega_i=v\oplus e_i)$ defined by $P:\omega_i\rightarrow x_i$, so that  the identification  in (\ref{DPL}) is given by $f^*(x_i)=f^*(P(\omega_i))$. Throughout the article, we will always consider $f^*(\omega_i)$ as $f^*(x_i)$ without mentioning the mapping $P$, using the fact that $S_f=v \oplus E$ is sorted with respect to some vector $v \in S_f$ and a lexicographically ordered set $E$. It appears to be difficult to capture  intrinsic properties of the mapping $P$ which would essentially establish the same results.
\end{rem}
\end{itemize}

\section{WHT of compositional form and plateaued functions}\label{sec:platechar}


In this section we first analyze the connection between the WHTs of functions $f,h_1,\ldots,h_k$ and $\mathfrak{f}$,
which is  later utilized to provide  various generic construction methods of bent and plateaued functions. In addition, a construction method of plateaued functions (of any amplitude) in terms of duals and Walsh supports is given.



\subsection{WHT of compositional form}\label{sec:formulas}

Let $\mathfrak{f}:\mathbb{F}^n_2 \rightarrow \mathbb{F}_2$ be an arbitrary Boolean function given in the $CF$-representation as
\begin{eqnarray}\label{F2}
\mathfrak{f}(x)=f(H(x))=f(h_1(x),\ldots,h_k(x)),
\end{eqnarray}
with the form $f:\mathbb{F}^k_2 \rightarrow \mathbb{F}_2$ and vectorial function $H=(h_1,\ldots,h_k):\mathbb{F}^n_2 \rightarrow \mathbb{F}^k_2$. Since in  (\ref{F2}) one can view $\mathfrak{f}(x)=f(H(x))$ as a composition of two vectorial functions, the result in \cite[Proposition 9.1]{CarletBoolean} implies that the WHT of $\mathfrak{f}$ is given by
\begin{eqnarray}\label{mainF}
W_\mathfrak{f}(u)&=&\sum_{x\in \mathbb{F}^n_2}(-1)^{f(h_1(x),\ldots,h_k(x))\oplus u\cdot x}=2^{-k}\sum_{\omega\in \mathbb{F}^k_2}W_f(\omega)W_{\omega\cdot (h_1,\ldots,h_k)}(u),\;\;\;u\in \mathbb{F}^n_2.
\end{eqnarray}
%
Hence, (\ref{mainF}) provides the relation between Walsh coefficients of functions $\mathfrak{f}$, $f$ and $h_1,\ldots,h_k.$
The non-uniqueness of this representation is evident from the fact that an arbitrary Boolean function $\mathfrak{f}$ can be written in any given form $f$ for some vectorial function $H$ which satisfies the sufficient conditions $H(supp(\mathfrak{f}))\subseteq supp(f)$ and $H(\mathbb{F}^n_2\backslash supp(\mathfrak{f}))\subseteq \mathbb{F}^k_2\backslash supp(f)$.

 Notice that, for the purpose of finding efficient  construction methods, plateaued forms (functions) are quite desirable due to a low number of nonzero linear combinations in $\omega\cdot (h_1,\ldots,h_k)$  and the fact that $W_f(\omega)=2^{\frac{k+s}{2}}(-1)^{f^*(\omega)}$, for $\omega\in S_f$,  implies that
\begin{eqnarray}\label{mainF2}
W_\mathfrak{f}(u)=2^{\frac{s-k}{2}}\sum_{\omega\in S_f}(-1)^{f^*(\omega)}W_{\omega\cdot (h_1,\ldots,h_k)}(u),\;\;\;u\in \mathbb{F}^n_2.
\end{eqnarray}
Clearly, when the form $f$ is a bent function, then in (\ref{mainF2}) we set $s=0$ and then $S_f=\mathbb{F}^k_2.$

The main purpose of this paper is to provide  generic secondary constructions of bent and plateaued functions using suitable forms and coordinate functions.
To make our method  highly efficient, in the following subsection we show how to construct a plateaued form by fixing its dual and Walsh support in advance. This will, in turn, give us the possibility of controlling the design process in terms of  the hardness and number of initial conditions used. 

%
%

\subsection{On plateaued functions and their duals}\label{sec:plateWS}

In this section, we provide a  method to specify the signs of a  Walsh spectrum so that by applying the inverse WHT to it a plateaued function is recovered. 
 At the same time, this approach  provides a new characterization of plateaued functions. We start with the following definition.
\begin{defi}\label{def:Sylv}
The \emph{Sylvester-Hadamard}  matrix of size $2^k \times 2^k$, is defined recursively as:
\begin{eqnarray*}\label{HM}
H_1=(1);\hskip 0.4cm H_2=\left(
                           \begin{array}{cc}
                             1 & 1 \\
                             1 & -1 \\
                           \end{array}
                         \right);\hskip 0.4cm H_{2^k}=\left(
      \begin{array}{cc}
        H_{2^{k-1}} & H_{2^{k-1}} \\
        H_{2^{k-1}} & -H_{2^{k-1}} \\
      \end{array}
    \right).
\end{eqnarray*}
The $i$-th row of $H_{2^k}$ its denoted by $H^{(i)}_{2^{k}}$. 
\end{defi}
\begin{lemma}\label{l1}
Let $S=\{\omega_{0},\ldots,\omega_{2^m-1}\}\subseteq \mathbb{F}^k_2$ be any  affine subspace of dimension $m\geq 2$ such that $S=v\oplus E$, for some lexicographically ordered linear subspace $E=\{e_0,\ldots,e_{2^m-1}\}\subseteq \mathbb{F}^k_2$ and $v\in S_f$, where $\omega_{i}=v\oplus e_i$ for $i\in [0,2^m-1]$. Then:
\begin{enumerate}[i)]
%
\item The lexicographic ordering of $E$ implies that for any fixed $i\in\{0,\ldots,m-1\}$ it holds that $e_j=e_{2^i}\oplus e_{j-2^i}$ for all $2^i\leq j\leq 2^{i+1}-1.$
\item For an arbitrary vector $u\in \mathbb{F}^k_2$ it holds that
\begin{eqnarray}\label{hrow}
((-1)^{u\cdot \omega_{0}},(-1)^{u\cdot \omega_{1}},\ldots,(-1)^{u\cdot \omega_{2^m-1}})=(-1)^{\varepsilon_u} H^{(r_u)}_{2^{m}},
\end{eqnarray}
for some $0\leq r_u\leq 2^m-1$ and $\varepsilon_u\in\mathbb{F}_2$. In addition,  $\{T_{\ell}:\ell\in \mathcal{L}_m\}\subseteq \{(u\cdot e_0,\ldots,u\cdot e_{2^m-1}):u\in \mathbb{F}^k_2\}$, which means that  $\mathcal{L}_m$ is contained in a multi-set of $m$-variable linear functions whose truth tables are $\{(u\cdot e_0,\ldots,u\cdot e_{2^m-1}):u\in \mathbb{F}^k_2\}$.


\end{enumerate}
%
\end{lemma}
\proof 
$i)$ Since $E \subseteq \F_2^k$ is a linear subspace and $\dim(E)=m$ any basis of $E$ can be represented as $m \times k$ binary matrix $G$ whose rows $\gamma_1,\ldots, \gamma_m$ are the basis vectors.
Transforming $G$ into the reduced row echelon form using Gauss-Jordan elimination,
we may assume that $G$ is given as $G=\left (\begin{array}{c} e^{(m)} \\ \vdots \\e^{(1)} \end{array} \right )$,
where $e^{(m)}>e^{(m-1)}>\ldots>e^{(1)}$ in terms of lexicographic ordering.  This follows from the definition of lexicographic ordering (having the most significant bit on the left) and the fact that the leading entry (the first non-zero entry from the left  also called a pivot) in any row $e^{(i)}$, for $i=m-1,\ldots,1$, is strictly to the left of the leading entry in the row  $e^{(i-1)}$. Notice also that every pivot is the only non-zero entry in its column, thus the columns containing pivots build the identity $m \times m$ matrix.

Now, we show that if $E$ is ordered lexicographically (in increasing order) as $E=\{e_0={\bf 0}_k, e_1, e_2, \ldots, e_{2^m-1}\}$ then the vector $e_i$ corresponds to $\sum_{j=0}^{m-1}i_je^{(m-j)}$, where $(i_0,\ldots,i_{m-1}) \in \F_2^m$ is the binary representation of integer $i \in [0,2^m-1]$. Since the pivots
 are the only non-zero entries in their columns
 then for any two distinct integers $i',i''\in[0,2^m-1]$ with the property that $(i'_0,\ldots,i'_{m-1})<(i''_0,\ldots,i''_{m-1})$ (that is when $i'<i''$), we clearly have that $\sum_{j=0}^{m-1}i'_je^{(m-j)}<\sum_{j=0}^{m-1}i''_je^{(m-j)}$.
This actually shows that $e_i=\sum_{j=0}^{m-1}i_je^{(m-j)}$.
Furthermore, for the lexicographically ordered space $\mathbb{F}^m_2$ we have  $x_{j}=x_{2^i}\oplus x_{j-2^i}$ ($x_j\in \mathbb{F}^m_2$) for arbitrary fixed $i\in[0,m-1]$ and all $2^i\leq j\leq 2^{i+1}-1$, which implies
\begin{eqnarray*}
e_j=x_j\cdot (e^{(m)},\ldots,e^{(1)})=x_{2^i}\cdot (e^{(m)},\ldots,e^{(1)})\oplus x_{j-2^i}\cdot (e^{(m)},\ldots,e^{(1)})=e_{2^i}\oplus e_{j-2^i},
\end{eqnarray*}
where $x_j\in \mathbb{F}^m_2$ is the binary representation of the integer $j$. Thus,  it necessarily holds that $e_{2^i}=e^{(i+1)}$, for all $i\in[0,m-1]$, i.e., the statement $i)$ holds.
$ii)$ Since by $i)$ for lexicographically ordered $E$ and $\mathbb{F}^m_2$ we have that $e_j\in E$ and $x_j\in \mathbb{F}^m_2$ satisfy the recursions $e_{j}=e_{2^i}\oplus e_{j-2^i}$ and  $x_{j}=x_{2^i}\oplus x_{j-2^i}$, it clearly implies that there exists a linear bijective mapping, say $\psi: \F_2^m \rightarrow E \subseteq \F_2^k$, which maps $x_{2^i}$ to $e_{2^i}$ (recall that $e_{2^i}=e^{(i+1)}$), i.e. $\psi(x_{2^i})=e_{2^i}$ $(i\in[0,m-1])$, and it preserves the ordering.
 More precisely, this mapping is defined by a binary matrix $A$ (of size $m \times k$) as $\psi(x_{2^i})=x_{2^i}A=e_{2^i}$, and since $wt(x_{2^i})=1$, we clearly have that the rows of $A$ are actually vectors $e_{2^i}$. Furthermore, the recursions $e_{j}=e_{2^i}\oplus e_{j-2^i}$ and  $x_{j}=x_{2^i}\oplus x_{j-2^i}$ trivially imply that $\psi(x_{j})=x_jA=e_j$ for all $j\in[0,2^m-1]$, which gives that $\psi$ preserves the lexicographic ordering.

On the other hand, for arbitrary $u\in \mathbb{F}^k_2$ and $j\in [0,2^m-1]$ it holds that $u\cdot e_j=u\cdot (x_jA)=uA^T\cdot x_j$. Since by \cite[Lemma 10]{SeberryX} we have that rows of the Sylvester-Hadamard matrix $H_{2^m}$ are sequences of linear functions in $m$ variables, it consequently holds that
 $$((-1)^{u\cdot \omega_0},\ldots,(-1)^{u\cdot \omega_{2^m-1}})=(-1)^{u\cdot v}((-1)^{c\cdot x_0},\ldots,(-1)^{c\cdot x_{2^m-1}})=(-1)^{\varepsilon_u}H^{(r_u)}_{2^m},$$
where $c=uA^T\in \mathbb{F}^m_2$, for some $r_u\in [0,2^m-1]$ and $\varepsilon_u\in \mathbb{F}_2$, i.e., relation (\ref{hrow}) holds. 

The second part  $ii)$ follows from the fact that $\dim(E)=m$ which means that $E$ contains $m$ linearly independent columns.  These columns are linear functions by (\ref{hrow}) when $E$ is represented as a binary $2^m \times k$ matrix whose rows are $e_i$, for $i=0,\ldots, 2^m-1$. Then,  $\{T_{\ell}:\ell\in \mathcal{L}_m\}\subseteq \{(u\cdot e_0,\ldots,u\cdot e_{2^m-1}):u\in \mathbb{F}^k_2\}$ since the latter set contains all linear combinations of $k$ columns of $E$.\qed
\begin{rem}
This lemma is crucial for establishing the properties of the so-called sequence profile and some other results  e.g. Lemma \ref{lema:delta} and Theorem \ref{exEth1}.
\end{rem}

Let $f$ be an arbitrary $s$-plateaued function defined on $\mathbb{F}^k_2$ and let $S_f=\{\omega_{0},\ldots,\omega_{2^{k-s}-1}\}\subseteq \mathbb{F}^k_2$ be its Walsh support written as $S_f=v\oplus E$, for some $v\in S_f$ and a lexicographically ordered set $E\subseteq \mathbb{F}^k_2$ ($\textbf{0}_k\in E$). The \emph{sequence profile} of $S_f$,
 which is a multi-set of $2^k$ sequences of length $2^{k-s}$ induced by $S_f$, is defined as
\begin{eqnarray*}\label{RSf}
\Phi^{(v,E)}_{f}=\{\phi_u:\mathbb{F}^{k-s}_2\rightarrow \mathbb{F}_2\;:\; \chi_{\phi_u}=((-1)^{u\cdot \omega_{0}},(-1)^{u\cdot \omega_{1}},\ldots,(-1)^{u\cdot \omega_{2^{k-s}-1}}),\;\omega_i\in S_f,\; u\in \mathbb{F}^{k}_2\}.
\end{eqnarray*}
For convenience, the sequence profile of $S_f$ throughout the article will be denoted by $\Phi_f$ instead of $\Phi^{(v,E)}_{f}$, although it is clear that it depends on the choice of $v \in S_f$ when representing $S_f=v\oplus E.$ Denoting by $b_1,\ldots,b_k$ the canonical basis of $\mathbb{F}^k_2$ ($b_i$ contains the non-zero coordinate at the $i$-th position), it is clear that $\Phi_{f}$ is spanned by the functions $\phi_{b_1},\ldots,\phi_{b_k}$, i.e., $\Phi_{f}=\langle \phi_{b_1},\ldots,\phi_{b_k}\rangle$. The following result provides a simple characterization of plateaued functions in terms of their dual and Walsh support.
\begin{theo}\label{theo:plateH}
Let  $S_f=v\oplus E=\{\omega_0,\ldots,\omega_{2^{k-s}-1}\} \subset \FB^k$, for some $v \in \mathbb{F}^k_2$ and subset $E=\{e_0,e_1, \ldots, e_{2^{k-s}-1}\}\subset \mathbb{F}^k_2$. For a function $f^*:\FB^{k-s}\rightarrow \mathbb{F}_2$ with $wt(f^*)=2^{k-s-1}\pm 2^{\frac{k-s}{2}-1}$, let the Walsh spectrum of $f$ be defined (by identifying $x_i \in \FB^{k-s}$ and $e_i \in E$) as
\begin{eqnarray}\label{eq:walsh}
W_f(u)= \left \{  \begin{array}{ll}
  2^{\frac{k+s}{2}}(-1)^{f^*(x_i)} & \textnormal{ for } u= v\oplus e_i \in S_f,\\
 0 & u \not \in S_f. \\
\end{array}  \right.
\end{eqnarray}
Then:
\begin{enumerate}[i)]
\item $f$ is an $s$-plateaued function if and only if $f^*$ is at bent distance to $\Phi_f$ defined by (\ref{RSf}), that is, for any $\phi_u\in \Phi_{f}$ it holds that $d_H(f^*,\phi_u)=2^{k-s-1}\pm 2^{\frac{k-s}{2}-1}$.
\item If $E\subset \mathbb{F}^k_2$ is a linear subspace such that (\ref{hrow}) holds, then $f$ is an $s$-plateaued function if and only if $f^*$  is a bent function on $\mathbb{F}^{k-s}_2$.
\end{enumerate}
\end{theo}
\proof $i)$ By the inverse WHT (relation (\ref{WHT})), at any $u\in \mathbb{F}^k_2$, we have
$$(-1)^{f(u)}=2^{-k}\sum_{\omega\in \mathbb{F}^k_2}W_f(\omega)(-1)^{\omega\cdot u}=2^{-k}\cdot 2^{\frac{k+s}{2}}\sum_{\omega\in S_f}(-1)^{f^*(\omega)\oplus\omega\cdot u}=2^{\frac{s-k}{2}}\chi_{f^*}\cdot \chi_{\phi_u},$$
since $f$ is an $s$-plateaued function. Equivalently, it holds that $\chi_{f^*}\cdot \chi_{\phi_u}=(-1)^{f(u)}2^{\frac{k-s}{2}}$, which means that $f^*$ is at bent distance to  $\phi_u\in \Phi_{f}$ ($u\in \mathbb{F}^k_2$ is arbitrary).

$ii)$ We have that $f$ is an $s$-plateaued function if and only if it holds that
$$\sum_{e\in E}(-1)^{f^*(v\oplus e)\oplus u\cdot e}=2^{\frac{k-s}{2}}(-1)^{f(u)\oplus u\cdot v}.$$
Since by Lemma \ref{l1} we have that $\{((-1)^{u\cdot e_0},\ldots,(-1)^{u\cdot e_{2^{k-s}-1}}):u\in \mathbb{F}^k_2\}$ contains sequences of all linear functions (in $k-s$ variables), we have that $f^*$ (defined as $f^*(v\oplus e_i)\leftrightsquigarrow f^*(x_i)$) is at bent distance to the set of all linear functions, i.e., $f^*$ is bent.\qed
\begin{rem}
From \cite{Ayca2} we have that Walsh support of a so-called partially bent function is a coset of the orthogonal complement of the space of its linear structures $\Lambda$, where $\Lambda=\{\alpha\in \mathbb{F}^k_2:f(x\oplus \alpha)\oplus f(x)=const.,\;\forall x\in \mathbb{F}^k_2\}$ for $f\in \mathcal{B}_k$. Thus, Theorem \ref{theo:plateH}-$(ii)$ concerns partially bent functions in terms of their duals and Walsh supports.
\end{rem}
The importance of Theorem \ref{theo:plateH} lies in the fact that the design of plateaued functions, in difference to the vast majority of other constructions, is achieved using (\ref{eq:walsh}) so that the design is moved to  the Walsh  spectral domain rather than working   in the ANF domain.
The following two examples  illustrate the construction of plateaued functions by specifying a dual and Walsh support in relation (\ref{eq:walsh}).

\begin{ex}\label{plateEX}
Let us construct a $3$-plateaued function $f:\mathbb{F}^5_2\rightarrow \mathbb{F}_2$ $(k=5,$ $s=3)$. Since $\#S_f=2^{k-s}=2^{5-3}=4$, then we need to select $4$ vectors from $\mathbb{F}^5_2$ that form a 2-dimensional affine subspace that constitute the Walsh support of $f$. In addition, we need a bent function $f^*$ in two variables. For instance, we can specify $S_f=\{\omega_{0},\omega_{1},\omega_{2},\omega_{3}\}\subseteq \mathbb{F}^5_2$ and $f^*$ as
\begin{eqnarray*}\label{Sff}
S_f=\left(
        \begin{array}{c}
          \omega_{0} \\
           \omega_{1} \\
           \omega_{2} \\
           \omega_{3} \\
        \end{array}
      \right)
=\left(
        \begin{array}{c}
          u_{[6]} \\
          u_{[13]} \\
          u_{[16]} \\
          u_{[27]} \\
        \end{array}
      \right)
=\left(
        \begin{array}{ccccc}
          0 & 0 & 1 & 1 & 0 \\
          0 & 1 & 1 & 0 & 1 \\
          1 & 0 & 0 & 0 & 0 \\
          1 & 1 & 0 & 1 & 1 \\
        \end{array}
      \right)=(0,0,1,1,0)\oplus \left(
        \begin{array}{ccccc}
          0 & 0 & 0 & 0 & 0 \\
          0 & 1 & 0 & 1 & 1 \\
          1 & 0 & 1 & 1 & 0 \\
          1 & 1 & 1 & 0 & 1 \\
        \end{array}
      \right)
\end{eqnarray*}
and $\chi_{f^*}=(1,1,1,-1)$, which also implies that  $T_{f^*}=(0,0,0,1)$. Here, $u_{[d]}$ denotes a binary vector  whose integer representation is $d \in [0,2^k-1]$. Clearly, $f$ is $3$-plateaued if $W_f(u)\in\{0,\pm 2^{\frac{5+3}{2}}\}=\{0,\pm 2^{4}\}$. Thus, using $S_f$ and the dual $f^*$ on $\FB^2$, the Walsh spectrum  $W_f$ of  $f$ can be constructed   so that
\begin{eqnarray*}
W_f&=&(0,\ldots,0,W_f(u_{[6]}),0,\ldots,0,W_f(u_{[13]}),0,0,W_f(u_{[16]}),0,\ldots,0,W_f(u_{[27]}),0,\ldots,0)\\
&=&(0,\ldots,0,2^{4},0,\ldots,0,2^{4},0,0,2^{4},0,\ldots,0,-2^{4},0,\ldots,0).
\end{eqnarray*}
Notice that the signs of non-zero values agree (lexicographically) with  the sequence $\chi_{f^*}=(1,1,1,-1).$
Now,  using the inverse WHT (\ref{WHT}), we recover the truth table of $f$ whose ANF is
 $f(x_1,\ldots,x_5)=x_4 (x_2 \oplus x_5) \oplus x_1 (x_2 \oplus x_4 \oplus x_5) \oplus x_3 (1 \oplus x_2 \oplus x_4 \oplus x_5)$, so that $f$ is  3-plateaued.
\end{ex}
\begin{ex}\label{counterPL}
Using Theorem \ref{theo:plateH}-$(i)$, one may construct a plateaued form with non-affine Walsh support as follows. Let $S_f=\mathbb{F}^4_2\wr  T_g$, where $g:\mathbb{F}^4_2\rightarrow \mathbb{F}_2$ is defined as $g(x_1,x_2,x_3,x_4)=x_3x_4\oplus 1$, and let $\chi_{f^*}$ be a sequence of the function $(x_1,x_2)\cdot (x_3,x_4)$. The last  column of $S_f$, hence  $T_g$, corresponds to the transpose of the truth table of  $g$, and $g$ is at bent distance to $f^*$, since $(x_1,x_2)\cdot (x_3,x_4)\oplus x_3x_4\oplus 1$ is a bent function in the $\mathcal{MM}$ class. The ANF of $f$ is given by $f(x_1,\ldots,x_5)=x_1x_2x_5\oplus x_1x_3\oplus x_2x_4\oplus x_5$, where $x_i \in \FB$.
%
 %
One can verify that $f$ is a $1$-plateaued (semi-bent) function, with the Walsh support which is not an affine subspace (i.e., $f$ is not a partially bent function).
\end{ex}

\section{Secondary constructions using disjoint variables}\label{sec:sepvar}

In this section we employ the compositional representation $\mathfrak{f}=f(h_1,\ldots,h_k)$ for the purpose of deriving new secondary constructions of bent/plateaued functions on larger variable spaces.
Using suitable linear coordinate functions,  we provide  several generalizations of the secondary construction method due to Rothaus \cite{Rot} (Section \ref{sec:linear}) and
of the  indirect sum of Carlet \cite{CarletRESB}. Thereby, we solve Open Problem 13 posed in \cite[Section 4.5]{CarletOP} regarding a generalization of indirect sums.

\subsection{Generalization of Rothaus' method - using  linear coordinate functions}\label{sec:linear}

The generalizations of the Rothaus construction (which we recall below), in terms of an increased number of initial functions with disjoint variables and stronger conditions (the linearity of dual function "$*$"), were  given in \cite[Corollary 1,2,3]{SecEnf}. Using the composite representation, we provide a generalization of Rothaus construction which appears to be more efficient than \cite{SecEnf} since it does not require the conditions related to the  linearity of "$*$" at all.
We start by recalling the secondary construction due to Rothaus.
\begin{theo}\cite{Rot}\label{rotth}
Let $(x,y_1,y_2)\in \mathbb{F}^k_2\times \mathbb{F}_2\times \mathbb{F}_2,$ and let  $a(x),b(x),c(x)$ and $a(x)\oplus b(x)\oplus c(x)$ be bent functions on $ \mathbb{F}^k_2$. Then
\begin{equation}\label{eq:rothoriginal}
\mathfrak{f}(x,y_1,y_2)=a(x)b(x)\oplus a(x)c(x)\oplus b(x)c(x)\oplus(a(x)\oplus b(x))y_2
\oplus(a(x)\oplus c(x))y_1\oplus y_1y_2
\end{equation}
is a bent function.
\end{theo}
%
In the following example we analyze the Rothaus construction in terms of composite representation.
\begin{ex}\label{rotex}
The function $\mathfrak{f}:\mathbb{F}^k_2\times \mathbb{F}_2\times \mathbb{F}_2\rightarrow \mathbb{F}_2$ in Theorem \ref{rotth} has the form $f$ given by: $$f(x_1,\ldots,x_5)=x_1x_2\oplus x_1x_3\oplus x_2x_3\oplus (x_1\oplus x_2)x_5\oplus (x_1\oplus x_3)x_4 \oplus x_4x_5,$$
where symbolically $x_1\leftrightarrow a(x),$ $x_2\leftrightarrow b(x),$ $x_3\leftrightarrow c(x),$ $x_{4}\leftrightarrow \ell_1(y_1,y_2)=y_1$ and $x_5\leftrightarrow \ell_2(y_1,y_2)=y_2.$ That is, $\mathfrak{f}(x,y_1,y_2)=f(a(x),b(x),c(x),y_1,y_2)$ where $x \in \FB^k$ and $y_i \in \FB$. The form $f$ is a $3$-plateaued function $(W_f(u)\in\{0,\pm 2^4\}$, $u\in \mathbb{F}^5_2)$ whose Walsh support is
\begin{eqnarray}\label{S_fM}
S_f=\left(
  \begin{array}{ccccc}
    1 & 0 & 0 & 0 & 0 \\
    0 & 1 & 0 & 1 & 0 \\
    0 & 0 & 1 & 0 & 1 \\
    1 & 1 & 1 & 1 & 1 \\
  \end{array}
\right)=S_1\wr S_2,
\end{eqnarray}
where  $S_1=\{(1,0,0),(0,1,0),(0,0,1),(1,1,1)\}$ and $S_2=\{(0,0),$ $(1,0),(0,1),(1,1)\}.$ Also, let the sequence of the dual $f^*$ be  given as $\chi_{f^*}=(1,1,1,-1).$ Using (\ref{mainF2}), it can be  verified that the WHT of $\mathfrak{f}$ at any $(u,v)\in \mathbb{F}^k_2\times \mathbb{F}^2_2$ is given by
\begin{eqnarray}\label{partition} \nonumber
W_\mathfrak{f}(u,v)
&=&
\left\{\begin{array}{cc}
                                                        2(-1)^{f^*(1,0,0,v)}W_a(u)=2W_a(u), & v=(0,0) \\
                                                        2(-1)^{f^*(0,1,0,v)}W_b(u)=2W_b(u), & v=(1,0) \\
                                                        2(-1)^{f^*(0,0,1,v)}W_c(u)=2W_c(u), & v=(0,1) \\
                                                        2(-1)^{f^*(1,1,1,v)}W_{a\oplus b\oplus c}(u)=-2W_{a\oplus b\oplus c}(u), & v=(1,1)
                                                      \end{array}
\right..
%
\end{eqnarray}
%
\end{ex}
From the above computation,
the bentness of $\mathfrak{f}$ is only governed by the  bentness of functions $\omega'\cdot (a,b,c)$ for $\omega'\in S_1$, and it is not affected by $f^*$.
This  is actually a consequence of the fact that $S_f$ can be written as $S_f=S_1\wr S_2$, where $S_2$ is equal to the (whole) vector space $\mathbb{F}^2_2$ of size $\#S_f=\#\mathbb{F}^2_2=2^2$, and the fact that the linear coordinate functions $\ell_1(y)=y_1$ and $\ell_2(y)=y_2$ are placed exactly at the coordinates (or variables) of $f$ which correspond to the set $S_2=\mathbb{F}^2_2$ (that is $x_4$ and $x_5$). Before we generalize this idea, we first fix the necessary notation.
For a function $f:\mathbb{F}^k_2\rightarrow \mathbb{F}_2$ let its Walsh support $S_f$ be written as $S_f=\Delta\wr \Theta$, where $\Delta$ is the set of the first $t$ ($<k$) coordinates of vectors $\omega\in S_f\subseteq \mathbb{F}^k_2$ and $\Theta$ is the set of the remaining $m=k-t$ coordinates of $\omega.$ More precisely, an arbitrary vector $\omega=(\omega_1,\ldots,\omega_t,\omega_{t+1},\ldots,\omega_k)\in S_f$  will be written as $\omega=(\delta,\theta)\in \Delta\wr \Theta=S_f$, where $\delta=(\omega_1,\ldots,\omega_t)\in \Delta$ and $\theta=(\omega_{t+1},\ldots,\omega_k)\in \Theta.$  Using relation (\ref{mainF}) one easily obtains the following result.

\begin{lemma}\label{rotlemma}
Let $H(x,y)=(h_1(x,y),\ldots,h_k(x,y)):\mathbb{F}^r_2\times\mathbb{F}^m_2\rightarrow \mathbb{F}^k_2$ be a vectorial function such that
\begin{eqnarray}\label{h_i}
\left\{\begin{array}{cc}
    h_i(x,y)=h_i(x), & i=1,\ldots,t,\;\; x\in \mathbb{F}^r_2, \\
    (h_{t+1}(x,y),\ldots,h_k(x,y))=(y_1,\ldots,y_m)=y\in  \mathbb{F}^m_2, & t+m= k.
  \end{array}\right.
\end{eqnarray}
Define $\mathfrak{f}:\mathbb{F}^r_2\times\mathbb{F}^m_2\rightarrow \mathbb{F}_2$  as $\mathfrak{f}(x,y)=f(H(x,y))$, where the Walsh support of $f:\mathbb{F}^k_2\rightarrow \mathbb{F}_2$ can be written as  $S_f=\Delta\wr \Theta$ with $\Theta=\mathbb{F}^m_2$ ($m=k-t$, $t \geq1$).
Then, for any  $(u,v)\in \mathbb{F}^r_2\times\mathbb{F}^m_2$ the WHT of  $\mathfrak{f}=f(h_1,\ldots,h_k)$ is given by
\begin{eqnarray*}
W_\mathfrak{f}(u,v)&=&2^{-t}\sum_{(\delta,v)\in S_f=\Delta\wr \Theta}W_{f}(\delta,v)W_{\delta\cdot (h_{1},\ldots,h_t)}(u), \;\; \delta \in \FB^t.
\end{eqnarray*}
\end{lemma}
%
\begin{rem}
 If  in Lemma \ref{rotlemma} we consider $\mathfrak{f}=a\oplus d(h_1,\ldots,h_k)$ with $a(x,y)=a(x)$, then in the WHT formula for $\mathfrak{f}$ instead of $W_{\delta\cdot (h_1,\ldots,h_t)}(u)$ we have $W_{a\oplus\delta\cdot (h_1,\ldots,h_t)}(u)$. Additionally, if we assume that $f:\mathbb{F}^k_2\rightarrow\mathbb{F}_2$ is an $s$-plateaued function (where $\#S_f=\Delta\wr \mathbb{F}^m_2=2^m=2^{k-s}$), then by relation (\ref{h_i}) with $t+m=k$ it necessarily holds that $t=s$.
\end{rem}
Employing  bent/plateaued coordinate functions $h_1,\ldots,h_s$ in $\mathfrak{f}=f(h_1,\ldots,h_s,y)$ we obtain the following result on the construction of bent/plateaued functions. 
Assuming that $\Theta$ is not a multi-set,
for any  $\omega=(\delta,\theta) \in S_f$, by $\vartheta_{\omega}: \Theta\rightarrow \Delta$ we denote the function which maps $\theta$ to $\delta$, hence $\vartheta_{\omega}(\theta)=\delta$.
\begin{theo}\label{P1}
Let $\mathfrak{f}:\mathbb{F}^r_2\times\mathbb{F}^m_2\rightarrow \mathbb{F}_2$ be given as $\mathfrak{f}(x,y)=f(H(x,y))=f(h_{1}(x),\ldots,h_s(x),y)$, where $f:\mathbb{F}^k_2\rightarrow \mathbb{F}_2$ is $s$-plateaued  and $H=(h_1,\ldots,h_k)$ is a vectorial function defined by (\ref{h_i}) ($t=s$). Assume that $S_f=\Delta\wr \Theta$  with $\Theta=\mathbb{F}^m_2$ and $\textbf{0}_s\not\in \Delta$ ($m\geq 2$ is even, $s+m=k$). Then:
 \begin{enumerate}[i)]
\item  If for every $\delta\in \Delta$ it holds that  $\delta\cdot (h_1,\ldots,h_s)$ is bent on $\mathbb{F}^r_2,$ then  $\mathfrak{f}$ is bent and its dual is given as $\mathfrak{f}^*(x,y)=f^*(\vartheta(y),y)\oplus (\vartheta(y)\cdot (h_1,\ldots,h_s))^*(x),$ $(x,y)\in \mathbb{F}^r_2\times \mathbb{F}^m_2.$

\item If for every $\delta\in \Delta$ it holds that  $\delta\cdot (h_1,\ldots,h_s)$ is $c$-plateaued  on $\mathbb{F}^r_2$ $(1\leq c\leq r-1)$, then $\mathfrak{f}$ is  $c$-plateaued.
\item If for every $\delta\in \Delta$ it holds that  $\delta\cdot (h_1,\ldots,h_s)$ is $c_{\delta}$-plateaued  with (possibly) different amplitudes $2^{\frac{r+c_{\delta}}{2}}$, then  $W_\mathfrak{f}(\omega) \in \{0,\pm 2^{\frac{r+m+c_{\delta}}{2}}:\delta\in \Delta\}$ and $\mathfrak{f}$ is not necessarily a three-valued spectra function.
\end{enumerate}
\end{theo}
The assumption $\Theta=\mathbb{F}^m_2$ implies that there  is a unique vector $\delta=\vartheta(v)\in \Delta$ such that $(\delta,v)\in S_f=\Delta\wr \Theta$.
Consequently, using $s=k-m$, we have
\begin{eqnarray*}
W_\mathfrak{f}(u,v)&=&2^{-s}W_{f}(\vartheta(v),v)W_{\vartheta(v)\cdot (h_{1},\ldots,h_s)}(u)=2^{-s+\frac{k+s}{2}+\frac{r}{2}}(-1)^{f^*(\vartheta(v),v)\oplus (\vartheta(v)\cdot (h_1,\ldots,h_s))^*(u)}\\
&=&2^{\frac{r+m+(m+s-k)}{2}}(-1)^{f^*(\vartheta(v),v)\oplus (\vartheta(v)\cdot (h_1,\ldots,h_s))^*(u)}=2^{\frac{r+m}{2}}(-1)^{\mathfrak{f}^*(u,v)},
\end{eqnarray*}
which means that $\mathfrak{f}$ is bent.\\
$ii)$ In this case,  we have
\[W_\mathfrak{f}(u,v)= \left \{ \begin{array}{ll}
0, & W_{\vartheta(v)\cdot (h_{1},\ldots,h_t)}(u)=0, \\
\pm 2^{\frac{r+m+c}{2}}, & W_{\vartheta(v)\cdot (h_{1},\ldots,h_t)}(u)=\pm 2^{\frac{r+c}{2}}.
\end{array} \right.
\]
In a similar way one shows $iii)$. \qed
\begin{rem}
Notice that  the restrictions $\mathfrak{f}_y(x)$ are not necessarily bent/plateaued functions, thus $\mathfrak{f}$ in Theorem \ref{P1} is in general not  a concatenation of bent/plateaued functions.
\end{rem}

Apparently,  the number of initial conditions in Theorem \ref{P1} (i.e., linear combinations $\delta\cdot (h_1,\ldots,h_s)$) depends on our choice of the set $\Delta$,   where  $\Delta$ must additionally satisfy the conditions of Theorem \ref{theo:plateH} so that the form $f$ can be constructed.
\subsection{Generalizations of Rothaus using reduced number of initial conditions}\label{subsec:genericRot}

 An efficient method of reducing the set of initial conditions in Theorem \ref{P1} is to select  $\Delta$ as a multi-set that contains many repeated vectors from $\mathbb{F}^s_2$.  This approach implies that, writing $\Delta$  as a matrix of size $2^m\times s$, the columns of $\Delta$ mainly correspond to  truth tables of affine/linear functions. In extreme case, all the columns of  $\Delta$ may correspond to affine functions   which then allows  us to choose any bent dual $f^*$ to construct the form $f$. This approach also implies that many linear combinations $\delta\cdot (h_1,\ldots,h_s)$ ($\delta\in \Delta$) are the same.
 The simplest method to achieve that the columns of  $\Delta$ are affine functions is to use suitable shifts of a linear subspace of small dimension.
\begin{lemma}\label{lema:delta}
Let $E=\{e_0,\ldots,e_3\}\subset \mathbb{F}^s_2$ be a linear subspace ($\dim(E)=2$) and $b_0,\ldots,$ $b_{2^{m-2}-1}\in E$ be arbitrary vectors (a multiset) such that  for $i\in[0,m-3]$ it holds that $b_j=b_{2^i}\oplus b_{j-2^i}$ $($for all $2^i\leq j \leq 2^{i+1}-1)$. 
Suppose that the set $\Delta$ is defined as $$\Delta=\{\delta_0,\ldots,\delta_{2^m-1}\}=v\oplus \{b_0\oplus E,\ldots,b_{2^{m-2}-1}\oplus E\},$$
where $v\in \mathbb{F}^s_2\setminus E$ and $b_j\oplus E=\{b_j\oplus e_0,\ldots,b_j\oplus e_3\}$. Then, $\Delta$ can be used in Theorem \ref{P1} to construct a plateaued form $f$ and $\{\delta\cdot (h_1,\ldots,h_s):\delta\in \Delta\}=\{(v\oplus e)\cdot (h_1,\ldots,h_s):e\in E\}$.
\end{lemma}
\proof Due to the construction of  $\Delta$ and recursion $b_j=b_{2^i}\oplus b_{j-2^i}$, it is not difficult to see that $\Delta$ satisfies the Sylvester-Hadamard recursion (Lemma \ref{l1}), which means that its columns (when written as a matrix of size $2^m\times s$) correspond to truth tables of affine/linear functions in $m$ variables. Since in Theorem \ref{P1} we have that $S_f=\Delta\wr \mathbb{F}^m_2$, then $S_f$ is an affine subspace and using an arbitrary bent dual $f^*:\mathbb{F}^m_2\rightarrow \mathbb{F}_2$ the form $f$ is easily constructed by Theorem \ref{theo:plateH}. Note that $v\in \mathbb{F}^s_2\setminus E$ implies that $\textbf{0}_s\not\in \Delta$. The second part follows from the fact that $b_j\in E$ and thus the statement holds.\qed
\begin{rem}
The dimension of  $E$ in Lemma \ref{lema:delta} directly affects the number of initial conditions in the set  $\{\delta\cdot (h_1,\ldots,h_s):\delta\in \Delta\}$, which gives us the possibility of controlling its cardinality. 
The form $f$ used in  Example \ref{rotex} is a special case of Lemma \ref{lema:delta}, since its Walsh support $S_f$ (given by (\ref{S_fM})) is equal to $\Delta\wr \mathbb{F}^2_2$ ($s=3,$ $m=2$, $s+m=k=5$), where $\Delta=\{(1,0,0),(0,1,0),(0,0,1),(1,1,1)\}$ is an affine subspace of $\mathbb{F}^3_2$ with $\dim(\Delta)=2$.
\end{rem}
The following result is implicitly based on an application of Lemma \ref{lema:delta} and it provides  an efficient secondary method that employs the same initial conditions as the original Rothaus construction.

\begin{theo}[Generalized Rothaus A]\label{theo:rotcompl}
Let $a,b,c \in \mathcal{B}_r$ be bent functions such that  $a\oplus b\oplus c$ is also bent. Then, $\mathfrak{f}(x,y):\mathbb{F}^r_2\times \mathbb{F}^4_2\rightarrow \F_2$, where $x \in \F_2^r$ and $y\in \F_2^4$, defined by
\begin{equation}\label{eq:ROT_A}
 \mathfrak{f}(x,y_1,\ldots,y_4)=b(x)(y_1 \oplus y_2) \oplus a(x) (1 \oplus y_1 \oplus y_3) \oplus (c(x) \oplus y_1) (y_2 \oplus y_3)\oplus (y_1 \oplus y_2) y_4,
\end{equation}
is a bent function.
\end{theo}
\begin{proof}
The proof follows from the design process of Lemma \ref{lema:delta} by specifying  a form $f:\mathbb{F}^7_2\rightarrow \mathbb{F}_2$ ($k=7$)  with parameters $m=4$ and $s=3$. Since $\#S_f=2^m=16$, we take $E=\{\textbf{0}_3,(1,1,0),(1,0,1),(0,1,1)\}$, $v=(1,0,0)$, and $\{b_0,\ldots,b_3\}=\{\textbf{0}_3,(1,0,1),(1,0,1),\textbf{0}_3\}$. The set $\Delta$ is given as
$\Delta=v\oplus\{E,(1,0,1)\oplus E,(1,0,1)\oplus E, E\}.$

Taking $\chi_{f^*}$ to be the sequence of a bent function $(x_1,x_2)\cdot (x_3,x_3 \oplus x_4) \oplus x_3x_4 \oplus x_3$ and using $S_f=\Delta\wr \mathbb{F}^4_2$, by Theorem \ref{theo:plateH}, we obtain the form
$$f(x_1,\ldots,x_7)=x_2 (x_4 \oplus x_5) \oplus x_1 (1 \oplus x_4 \oplus x_6) \oplus (x_3 \oplus x_4) (x_5 \oplus x_6)\oplus (x_4 \oplus x_5) x_7.$$
Now, defining $\mathfrak{f}(x,y)=f(a(x),b(x),c(x),y)$,
such that $a,b,c$ and $a\oplus b\oplus c$ are bent functions (see \cite{SecEnf} for specifying such bent functions),
we have that $\mathfrak{f}$ is bent and is given by (\ref{eq:ROT_A}). \qed
\end{proof}
A formal evidence that the above construction may generate  bent functions outside the $\mathcal{MM}$ class is given below in Example \ref{ex:4.7proof}. As discussed in \cite{SihemN}, showing that  a function $\mathfrak{f}:\mathbb{F}_2^m \times \mathbb{F}_2^m \rightarrow \mathbb{F}_2$ is outside $\mathcal{MM}$  corresponds to showing that
 the second order derivative of $\mathfrak{f}(x,y)$ defined (in general) as
$$ D_{\alpha}D_{\beta}\mathfrak{f}(x, y)=\mathfrak{f}(x,y)\oplus \mathfrak{f}(x\oplus \alpha',y \oplus \alpha'')\oplus \mathfrak{f}(x\oplus \beta',y\oplus \beta'')\oplus \mathfrak{f}(x\oplus \alpha' \oplus \alpha'', y \oplus \beta'\oplus \beta''), $$
where $\alpha=(\alpha', \alpha''),\beta=(\beta',\beta'') \in \mathbb{F}_2^m\times \mathbb{F}^m_2$ and $\alpha\neq \beta$, satisfies  $$D_{(\alpha',\textbf{0}_m)}D_{(\beta',\textbf{0}_m)}\vert _{x=\textbf{0}_m}\mathfrak{f}(x,y)\neq 0,$$
for some $\alpha',\beta' \in \mathbb{F}_2^{m}\setminus\{\textbf{0}_m\}$. Here, the notation $D_{(\alpha',\textbf{0}_m)}D_{(\beta',\textbf{0}_m)}\vert _{x=\textbf{0}_m}\mathfrak{f}(x,y)$ means that $x$ is set to be the all-zero vector $\textbf{0}_m$ after the second derivative $D_{(\alpha',\textbf{0}_m)}D_{(\beta',\textbf{0}_m)}\mathfrak{f}(x,y)$ has been computed.
\begin{ex} \label{ex:4.7proof}
Let $a,b,c:\mathbb{F}_2^2\times \mathbb{F}_2^2 \rightarrow \mathbb{F}_2, a(x_1,x_2)=x_1\cdot \pi_1(x_2) \oplus g_1(x_2), b(x_1,x_2)=x_1\cdot \pi_2(x_2) \oplus g_2(x_2), c(x_1,x_2)=x_1\cdot \pi_3(x_2) \oplus g_3(x_2)$, where $\pi_i(x_2)=x_2\oplus q_i$, for some arbitrary constants $q_i \in \mathbb{F}_{2}^2$ and arbitrary functions $g_i$ on $\FB^2$, for $i=1,2,3$. Since $\pi_1\oplus \pi_2\oplus \pi_3$ is a permutation, the sum of bent functions $a,b,c$ is again  a bent function in $\mathcal{MM}$. These initial functions are used to define $\mathfrak{f}(x_1,x_2,y)$ by means of (\ref{eq:ROT_A}), where $ \mathfrak{f}: \FB^2 \times \FB^2 \times \FB^4 \rightarrow \FB$.
Using the programming package Magma, it could be verified that  the second derivatives $D_{(\alpha',\textbf{0}_4)}$ $D_{(\beta',\textbf{0}_4)}\mathfrak{f}(x_1,x_2,y)$ do not vanish for many pairs $(\alpha',\beta')$ (with $\alpha' \neq \beta'$) when $x=(x_1,x_2)$ is set to be  $\textbf{0}_4$. This fact can also be  confirmed by computing $D_{(\alpha',\textbf{0}_4)}D_{(\beta',\textbf{0}_4)}\mathfrak{f}(x_1,x_2,y)$ and then specifying $\alpha'$ and $\beta'$ accordingly.  
This shows that this function is outside the $\mathcal{MM}$ class. Furthermore, since $a,b,c$ are quadratic functions then $\deg(\mathfrak{f})=3 < n/2$ (with $n=8$) and therefore $\mathfrak{f}$ does not belong to either $\mathcal{D}_0$ or to $\mathcal{PS}^{-}$.
\end{ex}
\begin{rem}
The construction in Theorem \ref{theo:rotcompl} is clearly inequivalent to the construction of Rothaus since it does not involve the products of initial bent functions. It can rather be viewed as a method to concatenate three bent functions $a,b$ and $c$  if the input variables $y_i$ are kept fixed.
\end{rem}
Another  generalization  of the Rothaus construction, which uses a form (defined on $\mathbb{F}^6_2$) that is not partially bent and  employs only two initial functions that depend on $x$ (whereas $y\in \mathbb{F}^4_2$), is given below.
\begin{theo}[Generalized Rothaus B]\label{afterrot}
Let  $a,b:\mathbb{F}^r_2\rightarrow \mathbb{F}_2$ ($r$ even) be two arbitrary bent functions.
 Then,  $\mathfrak{f}:\mathbb{F}^{r}_2 \times \mathbb{F}^4_2\rightarrow \mathbb{F}_2$  given by
\begin{equation} \label{eq:ROT_B}
\mathfrak{f}(x,y_1,\ldots,y_4)=b(x)\oplus (a(x)\oplus b(x))y_1y_2\oplus y_1y_3\oplus y_2y_4,
\end{equation}
is a bent function.
\end{theo}
\begin{proof}
By means of Lemma \ref{lema:delta},  we construct a form $f:\mathbb{F}^6_2\rightarrow \mathbb{F}_2$ using the Walsh support $S_f=\Delta\wr \mathbb{F}^2_2$ where the dual is given as $f^*(x_1,\ldots,x_4)=(x_1,x_2)\cdot (x_3,x_4)$ and $\Delta=T_g\wr T_{g\oplus 1}$ with $g(x_1,\ldots,x_4)=x_3x_4$ ($m=4$, $s=2$, and $k=6$). Notice that $f^*$ is at bent distance to $g$, since $f^*\oplus g$ belongs to the $\mathcal{MM}$ class of bent functions. The form $f$ is then given by
$$f(x_1,\ldots,x_6)=x_2 \oplus (x_1  \oplus x_2)x_3 x_4  \oplus x_3 x_5  \oplus x_4 x_6.$$
Since $\Delta=T_g\wr T_{g\oplus 1}$ is a multi-set which only contains the  vectors $(1,0)$ and $(0,1)$, then clearly $\delta\cdot (a,b)$ is always either equal to $a$ or $b$. Consequently, $\delta\cdot (a,b)$ is a bent function for all $\delta\in\Delta$ ($a,b$ are bent). The function $\mathfrak{f}=f(a,b,y_1,\ldots,y_4):\mathbb{F}^{r}_2\times \mathbb{F}^4_2\rightarrow \mathbb{F}_2$ is bent by Theorem \ref{P1}-$(i)$ and its ANF is given by (\ref{eq:ROT_B}). \qed
\end{proof}
Notice that there are no initial conditions on  bent functions $a$ and $b$ in this generalization, which corresponds to a special case of Rothaus when $b(x)=c(x)$. The latter identity   implies that the condition $a\oplus b \oplus c$ is bent is then automatically satisfied. Nevertheless, setting $b(x)=c(x)$ in Theorem \ref{rotth} then $\mathfrak{f}(x,y_1,y_2)=b(x)\oplus (a(x)\oplus b(x))(y_1\oplus y_2) \oplus y_1y_2$ is still bent for any bent functions $a$ and $b$.
However, setting  $y_3=y_4=0$ in Theorem \ref{afterrot} and taking, for instance, $b=x_1x_2$ and $a=x_1x_2\oplus x_1$ (which are bent on $\mathbb{F}^2_2$), then $\mathfrak{f}(x_1,x_2,y_1,y_2)=b(x)\oplus (a(x)\oplus b(x))y_1y_2\oplus y_1y_3\oplus y_2y_4=x_1x_2\oplus x_1y_1y_2$ is not a bent function. Since the restrictions of our bent functions are not bent in general, they  cannot trivially be obtained by any secondary construction of bent functions.
Another subtle but important difference in this context is that the restriction (by fixing $y_3=y_4=0$) involves the quadratic term $y_1y_2$ instead of $y_1\oplus y_2$ to multiply the sum $a(x)\oplus b(x)$, which for suitably chosen  $a$ and $b$ in (\ref{eq:ROT_B}) is affine inequivalent to  the bent function given by (\ref{eq:rothoriginal}) (when $b(x)=c(x)$).
\subsubsection{Generic concatenation methods} \label{subsec:generic_concat}
A more difficult  approach, with respect to the hardness of imposed conditions, is given by the following result. This approach essentially leads to a generalization of methods that concatenate initial bent functions to generate new ones.

\begin{theo}\label{P2}
Let $d:\mathbb{F}^{k}_2\rightarrow \mathbb{F}_2$ is a bent function ($k$ even), and $H=(h_1,\ldots,h_k)$ be a vectorial function defined by (\ref{h_i}) ($t+m=k$).  Assume that $a\oplus \langle h_{1},\ldots,h_t\rangle$ is an affine space of bent functions on $\mathbb{F}^r_2$.  Then, $\mathfrak{f}:\mathbb{F}^r_2\times \mathbb{F}^m_2\rightarrow \mathbb{F}_2$  given by $$\mathfrak{f}(x,y)=f(a(x),H(x,y))=a(x)\oplus  d(h_{1}(x),\ldots,h_t(x),y),$$
is a bent function
if for every $v\in \mathbb{F}^m_2$ the functions $\delta\rightarrow d^*(\delta,v)$ and $\delta\rightarrow (a\oplus \delta\cdot (h_1,\ldots,h_t))^*$ ($\delta\in \mathbb{F}^t_2$) are at bent distance.
\end{theo}
\proof By relation (\ref{mainF2}) and given assumptions ($t=k-m$), we have the following computation:
\begin{eqnarray*}
W_\mathfrak{f}(u,v)&=&2^{\frac{k}{2}-t}\sum_{(\delta,v)\in \mathbb{F}^k_2=\mathbb{F}^t_2\times \mathbb{F}^m_2}(-1)^{d^*(\delta,v)}W_{a\oplus \delta\cdot (h_1,\ldots,h_t)}(u)\\
&=&2^{\frac{k}{2}-t+\frac{r}{2}}\sum_{\delta\in \mathbb{F}^t_2}(-1)^{d^*(\delta,v)\oplus(a\oplus \delta\cdot (h_1,\ldots,h_t))^*(u)}=\pm 2^{\frac{k}{2}-t+\frac{r}{2}+\frac{t}{2}}=\pm 2^{\frac{r+m}{2}}.
\end{eqnarray*}\qed

An explicit construction method, also demonstrating that the sufficient conditions are not necessarily hard to satisfy, is given in the following corollary.
\begin{cor}[Bent concatenation]\label{cor:bent}
Let $f_1,f_2,f_3,f_4:\mathbb{F}^r_2\rightarrow \mathbb{F}_2$ be any four bent functions that satisfy $f^*_1\oplus f^*_2\oplus f^*_3\oplus f^*_4=1$, where $f_4=f_1\oplus f_2\oplus f_3.$
Then, $\mathfrak{f}:\mathbb{F}^r_2\times \mathbb{F}^2_2\rightarrow \mathbb{F}_2$ given by
\begin{equation} \label{eq:cor1} \mathfrak{f}(x,y_1,y_2)=f_1(x)\oplus y_1(f_1\oplus f_3)(x)\oplus y_2(f_1\oplus f_2)(x),
\end{equation}
is a bent function.
\end{cor}
\begin{proof}
Using the notation of Theorem \ref{P2}, let $d:\F_2^4 \rightarrow \F_2$, thus $k=4$, be given as $d(x_1,\ldots,x_4)=x_1 x_3 \oplus x_2 x_4.$ Then, $d$ is self-dual bent so that $d=d^*$ and furthermore $d$ is linear for any fixed $(x_1,x_2) \in \mathbb{F}^2_2$ or $(x_3,x_4) \in \mathbb{F}^2_2$. Now define, $a,h_1,h_2:\mathbb{F}^r_2\rightarrow \mathbb{F}_2$ so that  $a=f_1,$ $h_1=f_1\oplus f_3$ and $h_2=f_1\oplus f_2$. Then, after substitution, the function $\mathfrak{f}(x,y)=a(x)\oplus d(h_1(x),h_2(x),y):\mathbb{F}^r_2\times \mathbb{F}^2_2\rightarrow \mathbb{F}_2$ is given by (\ref{eq:cor1}).  The bentness of $\mathfrak{f}$ is due to the fact that $\delta\rightarrow (a(x)\oplus \delta\cdot (h_1(x),h_2(x)))^*$, $\delta\in \mathbb{F}^2_2$, is a bent function on $\mathbb{F}^2_2$ since $\sum_{\delta\in \mathbb{F}^2_2}(a\oplus \delta\cdot (h_1,h_2))^*=f^*_1\oplus f^*_2\oplus f^*_3\oplus f^*_4=1$, and the result follows by Theorem \ref{P2}. \qed
\end{proof}
\begin{rem}
The existence of quadruples of bent functions satisfying that $f^*_1\oplus f^*_2\oplus f^*_3\oplus f^*_4=1$ along with $f_4=f_1\oplus f_2\oplus f_3$ has been recently solved in \cite{Froben} for the purpose of constructing new families of bent functions using linear translators. This condition on duals is however different from $f^*_1\oplus f^*_2\oplus f^*_3\oplus f^*_4=0$ used by Mesnager \cite{Sihem}, see also Example \ref{dualcorex}.
\end{rem}
We notice that the restrictions of $\mathfrak{f}$ in Corollary \ref{cor:bent}, obtained by fixing $(y_1,y_2)\in \F_2^2$ are $f_1,f_2,f_3,f_4$, respectively (where $f_4=f_1\oplus f_2 \oplus f_3$).  On the other hand, the restrictions of original method of Rothaus (see eq. (\ref{eq:rothoriginal})) have more complicated expressions which have  quadratic terms that involve the initial bent functions. Indeed, renaming the bent functions $a,b,c$ in (\ref{eq:rothoriginal}) by $f_1,f_2,f_3$ the restriction of (\ref{eq:rothoriginal}) when $(y_1,y_2)=(0,0)$ corresponds to $f_1f_2\oplus f_1f_3 \oplus f_2f_3$ whereas in our case the same restriction is simply $f_1$. This observation indicates that, in general,  the two methods are  in  not equivalent.

\subsection{Generalization of the indirect sum}\label{sec:genindsum}

The so-called {\em indirect sum} has been introduced by Carlet in \cite{CarletRESB}. Based on the analysis of its form, we provide a generalized version of this method that employs larger sets of initial functions defined on disjoint variable spaces but without any initial conditions (except the bentness of initial functions). For this purpose we mainly use   plateaued forms with affine Walsh supports though we demonstrate the possibility to employ Walsh supports which are not affine. We start by recalling the indirect sum method.
\begin{cor}\cite{CarletRESB}\label{indsum}
Let $f_1$ and $f_2$ be bent functions on $\mathbb{F}^r_2$ ($r$ even) and $g_1$ and $g_2$ be bent functions defined on $\mathbb{F}^m_2.$ Then, $\mathfrak{f}:\mathbb{F}^r_2\times \mathbb{F}^m_2$
 defined as
\begin{eqnarray}\label{INDS}
\mathfrak{f}(x,y)=f_1(x)\oplus g_1(y)\oplus (f_1\oplus f_2)(x)(g_1\oplus g_2)(y),\;\;\;x\in \mathbb{F}^r_2, y\in \mathbb{F}^m_2,
\end{eqnarray}
is a bent function and its dual is obtained from $f^*_1$, $f^*_2$, $g^*_1$ and $g^*_2$ by the same formula
as $\mathfrak{f}$ is obtained from $f_1$, $f_2$, $g_1$ and $g_2.$
\end{cor}
To illustrate the idea of our approach, let us write the indirect sum given by (\ref{INDS}) as
$$\mathfrak{f}(x,y)=\xi(f_1(x),f_2(x),g_1(y),g_2(y))=f_1(x)\oplus g_1(y)\oplus (f_1\oplus f_2)(x)(g_1\oplus g_2)(y),$$
where $x\in \mathbb{F}^r_2, y\in \mathbb{F}^m_2$ and $\xi:\mathbb{F}^4_2\rightarrow \mathbb{F}_2$ is given as $\xi(x_1,x_2,x_3,x_4)=x_1\oplus x_3\oplus (x_1\oplus x_2)(x_3\oplus x_4).$ The function $\xi$ is semi-bent and its Walsh support is given by
\begin{eqnarray*}\label{S_eps}
S_{\xi}=\left(
          \begin{array}{cccc}
            0 & 1 & 0 & 1 \\
            0 & 1 & 1 & 0 \\
            1 & 0 & 0 & 1 \\
            1 & 0 & 1 & 0 \\
          \end{array}
        \right).
\end{eqnarray*}
The WHT of $\mathfrak{f}$ (cf. (\ref{mainF2}))  induces the linear combinations $\omega\cdot (f_1(x),f_2(x),g_1(y),g_2(y))$, where $\omega \in S_{\xi}.$ Due to the placement  of  ''ones'' in  $\omega \in S_{\xi}$ and the  fact that any $\omega$ is of  weight two,  we have that $\omega\cdot (f_1(x),f_2(x),g_1(y),g_2(y))$ is always equal to $f_i(x)\oplus g_j(y)$, for some $i,j\in\{1,2\}$.
Thus, these linear combinations
are never a sum of initial functions defined on the same variable space. This is the main reason why this indirect sum does not use  additional conditions apart from the  bentness of initial functions.

 Based on the above observation, our primary goal is to construct an affine Walsh support $S_{\xi} \subset \FB^k$
for a plateaued form $\xi$ so that $\omega\cdot (h_1,\ldots,h_k)$ is never equal to a sum of functions defined on the same  variable space, for any $\omega \in S_{\xi}$.
Furthermore, if we require that  $S_{\xi}$ is an affine subspace, then by Theorem \ref{theo:plateH} one can use any dual bent function $\xi^*$ in order to construct the form $\xi$.
The following technical result is useful for this purpose, where for convenience the truth table $T_{\ell}$ of an affine function $\ell$ is treated  as a column vector (as in the remainder of this section).
\begin{lemma}\label{methodL1}
Let $M=\{m_0,\ldots,m_{2^t-1}\}$ be a multi-set of vectors $m_i=(\ell(x_i),\ell(x_i)\oplus 1)$ for $i\in[0,2^t-1]$, where $\ell:\mathbb{F}^t_2 \rightarrow \mathbb{F}_2$ $(t \geq2)$ is an affine function, and $x_i\in \mathbb{F}^t_2$ (for lexicographically ordered $\mathbb{F}^t_2$). Then the set $M$, written as $M=T_{\ell}\wr T_{\ell\oplus 1}$, contains  vectors of weight one, and it holds that $\chi_M=((-1)^{u\cdot m_0},\ldots,(-1)^{u\cdot m_{2^t-1}})$  is a sequence of an affine function, for any $u\in \mathbb{F}^2_2$.
\end{lemma}
The application of  Lemma \ref{methodL1} for the case $t=2$ is given in the following example.
\begin{ex}\label{indsumth2}
Let us consider the sets $M_1,M_2,M_3$ given by
$$M_1=\left(
        \begin{array}{cc}
          0 & 1 \\
          0 & 1 \\
          1 & 0 \\
          1 & 0 \\
        \end{array}
      \right),\;\;M_2=\left(
        \begin{array}{cc}
          0 & 1 \\
          1 & 0 \\
          0 & 1 \\
          1 & 0 \\
        \end{array}
      \right),M_3=\left(
        \begin{array}{cc}
          1 & 0 \\
          1 & 0 \\
          0 & 1 \\
          0 & 1 \\
        \end{array}
      \right).
$$
The sets $M_1,M_2$ and $M_3$ contain complementary affine/linear functions as columns, and taking $S_{\xi}=M_1\wr M_2\wr M_3$ we get that $S_{\xi}=\{(0, 1, 0, 1, 1, 0), (0, 1, 1, 0, 1, 0), (1, 0, 0, 1, 0, 1),$ $(1, 0, 1, 0, 0, 1)\}\subset \mathbb{F}^6_2$ is an affine subspace whose sequence profile $\Phi_{\xi}$  lies in $\{\chi_g: g\in \mathcal{A}_2\}$.
\end{ex}
Using the ideas of Lemma \ref{methodL1} and taking a bent dual $\xi^*$ in four variables, one can easily construct suitable forms (by Theorem \ref{theo:plateH}) which then specifies a new indirect sum (cf. (\ref{eq:bentINDSUM_A})) defined on a set of four disjoint variables as follows.
\begin{theo}[Generalized indirect sum A]\label{newindsum}
Let  $H_i:\mathbb{F}^{r_i}_2\rightarrow \mathbb{F}^2_2$ ($r_i$ are even) be vectorial functions defined as $$H_1(x)=(f_1(x),f_2(x)), \; H_2(y)=(g_1(y),g_2(y)),\; H_3(z)=(l_1(z),l_2(z)), \; H_4(w)=(d_1(w),d_2(w)),$$
 where $f_i,g_i,l_i,d_i$ are bent functions, for $i=1,2$.
Then, $\mathfrak{f}:\mathbb{F}^{r_1}_2\times \mathbb{F}^{r_2}_2\times \mathbb{F}^{r_3}_2\times \mathbb{F}^{r_4}_2 \rightarrow \F_2$ given by
\begin{eqnarray}\label{eq:bentINDSUM_A}
\mathfrak{f}(x,y,z,w)&=& f_2(x)\oplus g_2(y)\oplus l_2(z)\oplus d_2(w)\oplus (g_1 \oplus g_2)(y)(f_1(x) \oplus f_2(x)\oplus d_1(w) \oplus d_2(w)) \nonumber \\
&&\oplus (l_1 \oplus l_2)(z)(f_1\oplus f_2)(x).
\end{eqnarray}
is a bent function. Furthermore, the dual of $\mathfrak{f}$ is obtained by the same formula
with the  coordinate functions $f_j,g_j,l_j,d_j$ replaced by $f^*_j,g^*_j$, $l^*_j$ and $d^*_j$.
\end{theo}
\begin{proof}
Let us first construct the form $\xi$ by Theorem \ref{theo:plateH}. For the linear functions $\ell_i=x_i$ on $\mathbb{F}^4_2$, we first define $M_i= T_{x_i}\wr T_{x_i\oplus 1}$, for $i\in[1,4]$.  Let now  $S_{\xi}=M_1\wr M_2\wr M_3\wr M_4\subset \mathbb{F}^8_2$ be the affine support of $\xi$ $(\#S_{\xi}=16)$,
and let the dual $\xi^*:\mathbb{F}^4_2\rightarrow \mathbb{F}_2$ be (a bent function) defined as $(x_1,x_2)\cdot (x_3,x_4)\oplus x_3x_4$ (in terms of relation (\ref{DPL})). By Theorem \ref{theo:plateH}-$(ii)$, a $4$-plateaued function $\xi$ on $\FB^8$ can be specified as
$$\xi(x_1,\ldots,x_8)=x_2\oplus x_4 \oplus x_6 \oplus x_8\oplus (x_3 \oplus x_4)(x_1 \oplus x_2\oplus x_7 \oplus x_8) \oplus (x_5 \oplus x_6)(x_1\oplus x_2).$$
Furthermore, the structure of $S_{\xi}$ implies that for every $\omega\in S_{\xi}$ it holds that $(\omega\cdot (H_1,H_2,H_3,H_4))^*$ $=\omega\cdot (H^*_1,H^*_2,H^*_3,H^*_4)$, where for instance $H^*_1=(f^*_1,f^*_2)$. Consequently, using the fact that $f_j,g_j,l_j$ and $d_j$ are bent, then for any $u=(u^{(1)},u^{(2)},u^{(3)},u^{(4)})\in \mathbb{F}^{r_1}_2\times \mathbb{F}^{r_2}_2\times \mathbb{F}^{r_3}_2\times \mathbb{F}^{r_4}_2$ we have that the WHT of $\mathfrak{f}(x,y,z,w)=\xi(H_1(x),H_2(y),H_3(z),H_4(w))$ is given as
\begin{eqnarray*}
W_\mathfrak{f}(u)&=&2^{-8}\sum_{\omega\in S_{\xi}}W_{\xi}(\omega)W_{\omega\cdot (H_1,H_2,H_3,H_4)}(u^{(1)},\ldots,u^{(4)})\\
&=&2^{-8+\frac{8+4}{2}+\frac{r_1+r_2+r_3+r_4}{2}}\sum_{\omega\in S_{\xi}}(-1)^{\xi^*(\omega)\oplus \omega\cdot (H^*_1,H^*_2,H^*_3,H^*_4)(u^{(1)},\ldots,u^{(4)})}\\
&\stackrel{(\ref{WHT})}{=}& 2^{\frac{n}{2}}(-1)^{\xi(H^*_1(u^{(1)}),H^*_2(u^{(2)}),H^*_3(u^{(3)}),H^*_4(u^{(4)}))}=2^{\frac{n}{2}}(-1)^{\mathfrak{f}^*(u)},
\end{eqnarray*}
thus $\mathfrak{f}$ is bent.\qed
%
\end{proof}
\begin{rem}
This method indeed generalizes the indirect sum method which is a special case obtained by removing  all the functions defined on $z$ and $w$ (by setting $l_i=d_i=0$). Indeed, in this case one obtains $\mathfrak{f}(x,y)=f_2(x)\oplus g_2(y)\oplus (g_1 \oplus g_2)(y)(f_1 \oplus f_2)(x)$ which is the indirect sum. On the other hand, setting $f_1=f_2$, the above construction  yields
$$[f_2(x)\oplus l_2(z)]\oplus [g_2(y)\oplus d_2(w)\oplus (g_1\oplus g_2)(y)(d_1\oplus d_2)(w)]$$
and due to the separation of variables in the rectangle brackets, this function is actually obtained by a direct sum method. Furthermore, if $f_2(x)=l_2(z)$, for $x=z$, then we only have the right bracket which is the indirect sum. This means that our construction contains both the direct and indirect sum method as special cases, whereas in the case $f_1\neq f_2$ it generates bent functions which in general cannot be obtained  either with direct or with indirect sum.
\end{rem}

Notice that   $S_{\xi}$ given in Example \ref{indsumth2} can be decomposed into two affine subspaces in $\mathbb{F}^3_2$, namely into the affine subspace $\{(0,0,1),(0,1,1),(1,0,0),(1,1,0)\}$, (taking odd numbered columns) and its complement (formed by even numbered columns). Alternatively, given any affine subspace $S \subseteq \FB^k$ of dimension $t$ one can form an affine subspace of the same dimension over $\FB^{2k}$ by  adding to each column of $S$ its binary complement.
This  observation  can be formalized as follows.
\begin{lemma}\label{methodL2}
Let $S'=T_{\ell_1}\wr \cdots \wr T_{\ell_k}\subseteq \mathbb{F}^k_2$ be an affine subspace ($\dim(S')=t$, $2\leq t\leq k$, $t$ is even), for some affine/linear functions $\ell_i:\mathbb{F}^t_2\rightarrow \mathbb{F}_2$. Then,  $S=M_1\wr\cdots\wr M_{k}$, with $M_i=T_{\ell_i}\wr T_{\ell_i\oplus 1}$  $(i\in[1,k])$, is  an affine subspace of $\mathbb{F}^{2k}_2$.
\end{lemma}
Notice that for an $s$-plateaued function on $\FB^n$,  for $n$ even or odd, the cardinality of its Walsh support is $2^{n-s}$ where $n$ and  $s$ are of the same parity, thus $n-s$ must be even.

To construct other indirect sum methods that use more disjoint variables, say $x^{(1)},\ldots,x^{(k)}$ (where $x^{(i)} \in \mathbb{F}_2^{r_i}$), we simply define coordinate functions $h_i$ on disjoint variable spaces such that a single variable corresponds to a multi-set $M_i$ over $\FB^2$ (which is composed out of two complementary truth tables of some linear/affine functions). This ensures that any two coordinate functions which depend on the same variable are not simultaneously present in $\omega \cdot (h_1,\ldots,h_k)$. The main steps of the design process are given below.

\begin{cons}[Indirect sum using $k>2$ disjoint variables - design steps]\label{genindsum} \mbox{}

\begin{itemize} \item For $i=1,\ldots,k$, where $k>2$, define $H_i(x^{(i)})=(f_{i,1}(x^{(i)}),f_{i,2}(x^{(i)})):\mathbb{F}^{r_i}_2\rightarrow \mathbb{F}^{2}_2$ such that $f_{i,1},f_{i,2}$ are bent, where $r_i$ is even.
\item Construct a plateaued form $\xi:\mathbb{F}^{m}_2\rightarrow \mathbb{F}_2$ by means of Theorem \ref{theo:plateH}-$(ii)$, with $m=2k$, whose (affine) Walsh support is constructed by  Lemma \ref{methodL2} and given by
$S_{\xi}=M_1\wr \cdots \wr M_k \subset \mathbb{F}^m_2$, where $\#S_{\xi}=\#M_i=2^{t}=2^{m-s},$ $s\in[1,m-1]$) and using arbitrary bent dual $\xi^*:\mathbb{F}^{t}_2\rightarrow \mathbb{F}_2$.
\item Define a bent function $\mathfrak{f}:\mathbb{F}^{r_1}_2\times \ldots\times \mathbb{F}^{r_k}_2\rightarrow \mathbb{F}_2$
  $(n=r_1+\ldots+r_k$) as
$$\mathfrak{f}(x^{(1)},x^{(2)},\ldots,x^{(k)})=\xi(H_1(x^{(1)}),H_2(x^{(2)}),\ldots,H_k(x^{(k)})),\;\;\;\;x^{(i)}\in \mathbb{F}^{r_i}_2.$$
\item
The dual of $\mathfrak{f}$ is  $\mathfrak{f}^*(x^{(1)},\ldots,x^{(k)})=\xi(H^*_1(x^{(1)}),\ldots,H^*_k(x^{(k)})),$ where $H^*_i=(f^*_{i,1},f^*_{i,2})$.
\end{itemize}
\end{cons}
\proof Using the fact that for all $\omega\in S_{\xi}$ it holds that $(\omega\cdot (H_1,\ldots,H_k))^*=\omega\cdot (H^*_1,\ldots,H^*_k)$, one shows the bentness of $\mathfrak{f}$ in the same way as in Theorem \ref{newindsum}.\qed

One should remark  that increasing  the size of Walsh support $S_{\xi}$ in Construction \ref{genindsum} has as a consequence that both the number of variables of the corresponding form and the number of employed disjoint variables  are increased as well. Nevertheless, one can always decrease the number of disjoint variables and overall number of initial bent functions by using suitable linear coordinate functions as in Section \ref{sec:linear}. The following result implicitly combines  the ideas of Lemma \ref{methodL2} and Theorem \ref{P1} for this purpose.
\begin{theo}[Generalized indirect sum B]\label{newindsumm}
%
Let $f_1,f_2:\mathbb{F}^r_2\rightarrow \mathbb{F}_2$ ($r$ even) and $g_1,g_2:\mathbb{F}^m_2\rightarrow \mathbb{F}_2$ ($m$ even) be arbitrary bent functions.
Then,  $\mathfrak{f}:\mathbb{F}^r_2\times \mathbb{F}^m_2\times \mathbb{F}^2_2\rightarrow \mathbb{F}_2$ defined by  
\begin{equation}\label{eq:indsum_C}
\mathfrak{f}(x,y,z_1,z_2)=f_2(x)\oplus g_2(y)\oplus(g_1(y)\oplus g_2(y)\oplus z_2)(f_1(x)\oplus f_2(x)\oplus z_1),
\end{equation}
is a bent function.
\end{theo}
\begin{proof}
Let us consider the Walsh support $S_{\xi}$ given as $S_{\xi}=M_1\wr M_2\wr \mathbb{F}^2_2\subset \mathbb{F}^6_2$, where $M_1$ and $M_2$ are given as in Example \ref{indsumth2}. Using the bent dual $\xi^*(x_1,x_2)=x_1x_2$, by Theorem \ref{theo:plateH} we construct the $4$-plateaued form $\xi(x_1,\ldots,x_6)=x_2\oplus x_4\oplus (x_3\oplus x_4\oplus x_6)(x_1\oplus x_2 \oplus x_5).$
Then, by Theorem \ref{P1}-$(i)$, $\mathfrak{f}(x,y,z_1,z_2)=\xi(f_1(x),f_2(x),g_1(y),g_2(y),z_1,z_2)$ is bent since for all $\delta\in M_1\wr M_2$ we trivially have that $\delta\cdot (f_1,f_2,g_1,g_2)$ are bent functions on $\mathbb{F}^r_2\times \mathbb{F}^m_2.$ It is easily verified that  its ANF is given by (\ref{eq:indsum_C}). \qed
\end{proof}

Clearly, the restriction $\mathfrak{f}(x,y,0,0)$ corresponds to a bent function on $\mathbb{F}^r_2\times \mathbb{F}^m_2$ obtained by the indirect sum of Carlet. On the other hand, the remaining three restrictions (corresponding respectively to fixing $(z_1,z_2)$ to (1,0),(0,1) and (1,1)) are given as  $\mathfrak{f}(x,y,0,0) \oplus g_1 \oplus g_2$, $\mathfrak{f}(x,y,0,0) \oplus f_1 \oplus f_2$ and $\mathfrak{f}(x,y,0,0) \oplus g_1 \oplus g_2 \oplus f_1 \oplus f_2$. This implies that  $\mathfrak{f}(x,y,z_1,z_2)$ given by (\ref{eq:indsum_C}) cannot be viewed as a trivial extension (a concatenation of the form $h ||h||h||(1\oplus h)$) of the indirect sum of Carlet since its restrictions differ by certain sums of initial bent functions.

We conclude this section by providing yet another secondary construction which uses a  form whose Walsh support is not affine subspace.
 In this context, we show that Construction \ref{genindsum} can be modified so that it contains multi-sets $M_i$ whose columns are non-linear functions which are still  at bent distance to the dual $\xi^*$ in terms of Theorem \ref{theo:plateH}-$(i)$.
\begin{theo}[Generalized indirect sum C]\label{newindsuM}
Let $f_1,f_2:\mathbb{F}^r_2\rightarrow \mathbb{F}_2$ ($r$ even) and $g_1,g_2:\mathbb{F}^m_2\rightarrow \mathbb{F}_2$ ($m$ even) be arbitrary bent functions.
Then,  $\mathfrak{f}:\mathbb{F}^r_2\times \mathbb{F}^m_2\times \mathbb{F}^4_2\rightarrow \mathbb{F}_2$ defined by  
\begin{eqnarray}\label{eq:indsum_C2}\nonumber
\mathfrak{f}(x,y,z_1,\ldots,z_4)&=&f_2(x)\oplus g_2(y)\oplus z_1(g_1(y)\oplus g_2(y)\oplus z_2)(f_1\oplus f_2)(x)\\
&&\oplus z_1z_4(g_1\oplus g_2)(y)\oplus z_2z_4\oplus z_1z_3,
\end{eqnarray}
is a bent function.
\end{theo}
\begin{proof}
Let us consider the Walsh support $S_{\xi}$ given as $S_{\xi}=M_1\wr M_2\wr \mathbb{F}^4_2\subset \mathbb{F}^8_2$, where $M_1=T_{q_1}\wr T_{q_1\oplus 1}$ and $M_2=T_{q_2}\wr T_{q_2\oplus 1}$ with $q_1(x_1,\ldots,x_4)=x_3x_4$ and $q_2(x_1,\ldots,x_4)=x_2x_3$. Taking the bent function $\xi^*(x_1,\ldots,x_4)=(x_1,x_2)\cdot(x_3,x_4)$, one may verify that $\xi^*\oplus c_1q_1\oplus c_2q_2$ is bent on $\mathbb{F}^4_2$ for all $(c_1,c_2)\in \mathbb{F}^2_2$, and thus by Theorem \ref{theo:plateH} we obtain the $4$-plateaued form $\xi:\mathbb{F}^8_2\rightarrow \mathbb{F}_2$ given as
$$\xi(x_1,\ldots,x_8)=x_2\oplus x_4\oplus x_5(x_3\oplus x_4\oplus x_6)(x_1\oplus x_2)\oplus x_5x_8(x_3\oplus x_4)\oplus x_6x_8\oplus x_5x_7.$$
Then, by Theorem \ref{P1}-$(i)$, $\mathfrak{f}(x,y,z_1,\ldots,z_4)=\xi(f_1(x),f_2(x),g_1(y),g_2(y),z_1,\ldots,z_4)$ is bent since for all $\delta\in M_1\wr M_2$ it holds that $\delta\cdot (f_1,f_2,g_1,g_2)$ are bent functions on $\mathbb{F}^r_2\times \mathbb{F}^m_2.$ \qed
\end{proof}
\begin{rem} Once again, by  fixing $(z_1,\ldots,z_4)=(1,0,0,0)$, the restriction $\mathfrak{f}(x,y,1,0,0,0)=f_2(x)\oplus g_2(y)\oplus (f_1\oplus f_2)(x)(g_1\oplus g_2)(y)$ is actually the indirect sum. On the other hand, the restriction to a hyperplane  $z_1=0$ equals to $f_2(x)\oplus g_2(y)\oplus z_2z_4$ which is essentially the direct sum method.  The latter form corresponds to a semi-bent or bent function, if considered   on $\mathbb{F}^{r+m+4}_2$ or $\mathbb{F}^{r+m+2}_2$, respectively.
\end{rem}

\section{Bent functions without increasing the variable space}\label{sec:bent}

In this section, we focus on secondary constructions of bent functions without increasing the number of variables, thus $\mathfrak{f}=f(h_1,\ldots,h_k)$ will be defined on the same number of variables as  $h_1,\ldots,h_k$. In general, there are only a few secondary constructions of this type treated in the literature \cite{CarletRes,Sihem,Sihem2}.
  The difficulty of specifying new methods (without using functions on disjoint variable spaces) lies in the fact that the initial functions  (commonly) need to satisfy very strong conditions that  are in general related to linearity of  "$*$", see Section \ref{sec:dual}. Therefore, to slightly relax  these hard conditions, in Section \ref{sec:dual2} we provide some  explicit constructions of bent and plateaued functions by employing an indicator set  as the form $f$.

\subsection{Necessary and sufficient conditions using bent/plateaued form}\label{sec:dual}

Similarly to the multi-set of sequences $\Phi_f$, for the function $\mathfrak{f}=f(h_1,\ldots,h_k)$ one can define
$$\Phi_{f,h}=\{\varphi_u:\mathbb{F}^{k-s}_2\rightarrow \mathbb{F}_2:\chi_{\varphi_u}=((-1)^{(\omega_{0}\cdot (h_1,\ldots,h_k))^*(u)},\ldots,(-1)^{(\omega_{2^{k-s}-1}\cdot (h_1,\ldots,h_k))^*(u)}),\;u\in \mathbb{F}^n_2\},$$
where $\omega_{i}\in S_f=v\oplus E$ ($E$ ordered lexicographically containing $\textbf{0}_k$, $\#S_f =2^{k-s}$), which is also a multi-set called  {\em the sequence profile of  $\mathfrak{f}$}.  As in the case of $\Phi_f$, the set $\Phi_{f,h}$ depends on  $S_f$ and its representation as $S_f=v\oplus E$ ($v\in S_f$). Then, for an arbitrary $u\in \mathbb{F}^n_2$ we assign:
\begin{eqnarray}\label{W}
W_{S_f,h}(u)=(W_{\omega_{0}\cdot (h_1,\ldots,h_k)}(u),W_{\omega_1\cdot (h_1,\ldots,h_k)}(u),\ldots,W_{\omega_{2^{k-s}-1}\cdot (h_1,\ldots,h_k)}(u)).
\end{eqnarray}
\begin{rem}\label{phifh}
If for every $\omega\in S_f$ it holds that $(\omega\cdot (h_1,\ldots,h_k))^*=\omega\cdot(h^*_1,\ldots,h^*_k),$ then $\Phi_{f}$ contains the functions $\varphi_u\in \Phi_{f,h}$. In fact, $\Phi_{f}$ contains pairwise different functions from $\Phi_{f,h}$ if and only if $Im(h)=\mathbb{F}^k_2$, i.e.,   $h$ is surjective.
\end{rem}
Assuming that the coordinate functions $h_1,\ldots,h_k:\FB^n \rightarrow \FB$ are bent, we now give both necessary and sufficient conditions so that $\mathfrak{f}=f(h_1,\ldots,h_k)$ is bent. The proof is similar to the proof of Theorem \ref{theo:plateH}-$(i)$ and therefore omitted.
\begin{prop}\label{p1}
Let $\mathfrak{f}=f(h_1,\ldots,h_k),$ $\mathfrak{f}:\mathbb{F}^n_2 \rightarrow \mathbb{F}_2$, $n$ even, where  $f:\mathbb{F}^k_2 \rightarrow \mathbb{F}_2$ is  $s$-plateaued such that $\textbf{0}_k\not\in S_f$. In addition, assume that $\omega\cdot (h_1,\ldots,h_k)$ are bent functions on $\mathbb{F}^n_2$, for all $\omega\in S_f.$ Then, $\mathfrak{f}$ is a bent function if and only if  $d_H(f^*,\Phi_{f,h})=2^{k-s-1}\pm 2^{\frac{k-s}{2}-1}$.
\end{prop}
Using Proposition \ref{p1}, one can easily prove the following result.
\begin{cor}\label{cor:pf1}
Let $\mathfrak{f}:\mathbb{F}^n_2 \rightarrow \mathbb{F}_2$, $n$ even, be given as $\mathfrak{f}=f(a,h_1,\ldots,h_k)=a\oplus d(h_1,\ldots,h_k)$ so that $a\oplus \omega\cdot(h_1,\ldots,h_k)$ is bent on $\mathbb{F}^n_2$, for all $\omega\in S_d$.  If $d$ is  bent or plateaued  on $\mathbb{F}^k_2$, then $\mathfrak{f}$ is bent  if and only if $d^*$ is at bent distance to the function $\omega\rightarrow (a\oplus \omega\cdot(h_1,\ldots,h_k))^*(u)$ ($\omega\in S_d$), for all $u\in \mathbb{F}^n_2.$
\end{cor}
\proof Let us assume that $d$ is an $s$-plateaued function (for $s=0$ we have that $d$ is bent). For arbitrary $u\in \mathbb{F}^n_2$, the WHT of $\mathfrak{f}$ (by relation (\ref{mainF2})) is given as
\begin{eqnarray*}
W_{\mathfrak{f}}(u)=2^{\frac{s-k}{2}}\sum_{\omega\in S_d}(-1)^{d^*(\omega)}W_{a\oplus \omega\cdot (h_1,\ldots,h_k)}(u)=2^{\frac{s-k}{2}+\frac{n}{2}}\sum_{\omega\in S_d}(-1)^{d^*(\omega)\oplus (a\oplus \omega\cdot(h_1,\ldots,h_k))^*(u)}.
\end{eqnarray*}
Since $\#S_d=2^{k-s}$, then clearly $\mathfrak{f}$ is bent if and only if $d^*$ is at bent distance to $\omega\rightarrow (a\oplus \omega\cdot(h_1,\ldots,h_k))^*(u)$ ($\omega\in S_d$), i.e., the statement holds.\qed
In general, the  conditions in Corollary \ref{cor:pf1} are quite difficult to satisfy. To simplify these conditions, we assume the linearity  of "$*$" (i.e. the condition $(\omega\cdot (h_1,\ldots,h_k))^*=\omega\cdot (h^*_1,\ldots,h^*_k)$) which induces the following result.
\begin{theo}\label{p2}
Let $\mathfrak{f}:\mathbb{F}^n_2 \rightarrow \mathbb{F}_2$ ($n$ even) be given as $\mathfrak{f}=f(h_1,\ldots,h_k),$ where $f$ is  $s$-plateaued on $\FB^k$ and $\textbf{0}_k\not\in S_f$. Assume that for every $\omega\in S_f$ the functions  $\omega\cdot (h_1,\ldots,h_k)$ and $h_i$ are bent such  that $(\omega\cdot (h_1,\ldots,h_k))^*=\omega\cdot (h^*_1,\ldots,h^*_k)$. Then:
\begin{enumerate}[i)]
\item $\mathfrak{f}$ is bent  and its dual is given by $\mathfrak{f}^*=f(h^*_1,\ldots,h^*_k).$
\item If $h_1,\ldots,h_k$ are self-dual bent, i.e., $h_i=h^*_i$ for $i=1,\ldots,k$, then $\mathfrak{f}$ is  self-dual bent.
\end{enumerate}
\end{theo}
\proof Since by assumption  $(\omega\cdot (h_1(u),\ldots,h_k(u)))^*=\omega\cdot (h^*_1(u),\ldots,h^*_k(u))$ holds for every $\omega\in S_f$, we have that $W_{S_f,h}(u)=2^{\frac{n}{2}}\chi_{\varphi_{u}}$, where $\varphi_{u}\in \Phi_{f}$ ($u\in \mathbb{F}^n_2$ is arbitrary). By Proposition \ref{p1},  $\mathfrak{f}$ is bent.
%
The dual of $\mathfrak{f}$ can be easily derived from (\ref{WHT}).
%
 If $h_1,\ldots,h_k$ are self-dual bent, then $\mathfrak{f}^*=f(h^*_1,\ldots,h^*_k)=f(h_1,\ldots,h_k)=\mathfrak{f}$.\qed

In the following example, we recall and analyze Theorem 3 in \cite{Sihem} which is shown to employ a plateaued form and affine Walsh support. This result (\cite[Theorem 3]{Sihem})  is just a particular case of Corollary \ref{cor:pf1} and Theorem \ref{p2}, thus it implicitly uses the linearity of duals.
\begin{ex}\cite[Theorem 3]{CarletRes}\label{ex:sihem}
Let $n$ be an even integer. Let $f_1$, $f_2$ and $f_3$
be three pairwise distinct bent functions over $\mathbb{F}^n_2$ such that
$\psi = f_1 \oplus f_2 \oplus f_3$ is bent. Let $g$ be a Boolean function defined
by $g=f_1 f_2\oplus f_1 f_3\oplus f_2 f_3$. Then, $g$ is bent if and only if $f^*_1 \oplus f^*_2  \oplus f^*_3 \oplus \psi^* = 0.$
Furthermore, if $g$ is bent then its dual function $g^*$ is given by
$$g^*=f^*_1 f^*_2\oplus f^*_1 f^*_3\oplus f^*_2 f^*_3.$$

The form $f$ of  $g$  is a  quadratic function $f(x_1,x_2,x_3)=x_1x_2\oplus x_1x_3\oplus x_2x_3,$ i.e., $g$ can be written as $g(x)=f(f_1(x),f_2(x),f_3(x)).$ The form $f$ is a semi-bent  function, with Walsh support $S_f=\{(1,0,0),(0,1,0),(0,0,1),(1,1,1)\}$ and the dual function $\chi_{f^*}=(1,1,1,-1).$ It is easy to verify that the functions $\omega\cdot (f_1,f_2,f_3)$, $\omega\in S_f,$ satisfy the equality  $W_{S_f,h}(u)=\pm 2^{\frac{n}{2}}H^{(r)}_4$ (for some $0\leq r\leq 3$, $u\in \mathbb{F}^n_2$), due to the structure of  $S_f.$ In addition, the dual of  $g$ is given by $g^*=f(f^*_1,f^*_2,f^*_3)$ (due to Theorem \ref{p2}-$i$)).
\end{ex}
\begin{rem}
Note that based on \cite[Theorem 4]{Sihem} (or \cite[Theorem 3]{CarletRes}) and \cite[Corollary 5]{Sihem},  Mesnager \cite{Sihem,Sihem2}  derived several new infinite classes of bent functions and specified their duals. In this context, we emphasize that our framework allows us to specify many infinite  families of bent functions   similar to those in \cite{Sihem,Sihem2} through different forms and  specification of the corresponding initial functions.
\end{rem}
The condition that $\omega\cdot (h_1,\ldots,h_k)$ is bent for all $\omega \in S_f$ in Theorem \ref{p2} seems to be necessary but the condition on the duals can be slightly relaxed.
\begin{prop}\label{dualcor}
Let $\mathfrak{f}:\mathbb{F}^n_2 \rightarrow \mathbb{F}_2$ ($n$ even) be given as $\mathfrak{f}=f(h_1,\ldots,h_k),$ where $f$ is an $s$-plateaued function such that $\textbf{0}_k\not\in S_f$. In addition, let $\omega\cdot (h_1,\ldots,h_k)$ be bent on $\mathbb{F}^n_2$, for all $\omega\in S_f$. If there exist functions $h'_1,\ldots,h'_k:\mathbb{F}^n_2\rightarrow \mathbb{F}_2$ (not necessarily bent) such that $(\omega\cdot (h_1,\ldots,h_k))^*=\omega\cdot (h'_1,\ldots,h'_k)$ for all $\omega\in S_f$, then $\mathfrak{f}$ is bent and $\mathfrak{f}^*=f(h'_1,\ldots,h'_k).$
\end{prop}
\proof Using the assumption that for all $\omega\in S_f$, the function $\omega\cdot (h_1,\ldots,h_k)$ is bent on $\mathbb{F}^n_2$ and that $(\omega\cdot (h_1,\ldots,h_k))^*=\omega\cdot (h'_1,\ldots,h'_k)$ holds, then for arbitrary $u\in \mathbb{F}^n_2$ we have
$$W_{\mathfrak{f}}(u)\stackrel{(\ref{mainF})}{=}2^{\frac{n}{2}}\cdot 2^{-k}\sum_{\omega\in S_f}W_f(\omega)(-1)^{(\omega\cdot (h_1(u),\ldots,h_k(u)))^*}\stackrel{(\ref{WHT})}{=}2^{\frac{n}{2}}(-1)^{f(h'_1(u),\ldots,h'_k(u))},$$
which means that $\mathfrak{f}$ is bent.\qed

To illustrate the efficiency of Proposition \ref{dualcor}, we give the following example.
\begin{ex}\label{dualcorex}
Consider the quadratic form  $f(x_1,x_2,x_3,x_4)=x_3 (x_2 \oplus x_4) \oplus x_1 (x_2 \oplus x_3 \oplus x_4)$, whose Walsh support is given as $S_f=\{(1,0,0,0),(0,1,0,1),(0,0,1,0),(1,1,1,1)\}$. We define  $h_1,h_2,h_3,h_4:\mathbb{F}^{\frac{n}{2}}_2\times \mathbb{F}^{\frac{n}{2}}_2\rightarrow \mathbb{F}_2$ as $$h_1(x,y)=x\cdot \pi(y)\oplus g_1(y), \; h_2(x,y)=x\cdot \pi(y), \; h_3(x,y)=x\cdot \varphi(y),\; h_4(x,y)=g_2(y),$$ where $g_1,g_2\in \mathcal{B}_{n/2}$ are arbitrary and $\pi, \varphi$ are permutations over $\mathbb{F}^{\frac{n}{2}}_2$. Notice that $h_4$ is arbitrary thus not necessarily bent. Let $\lambda=g_1\oplus g_2\in \mathcal{B}_{n/2}$, $\eta=g_1g_2\in \mathcal{B}_{n/2}$ and assume  that $\pi$ and $\varphi$ satisfy $\lambda(\pi^{-1}(y))=\lambda(\varphi^{-1}(y))$ \footnote{For $A=\{y\in \mathbb{F}^{n/2}_2:\;\pi^{-1}(y)\in supp(\lambda)\}$, it holds that $\varphi^{-1}(A)=\pi^{-1}(A)$, and similarly for $\mathbb{F}^{n/2}_2\setminus A.$}, for all $y\in \mathbb{F}^{n/2}_2$. Then for $(h'_1,h'_2,h'_3,h'_4)(x,y)=(h^*_1(x,y),h^*_2(x,y),h^*_3(x,y),g_2(\pi^{-1}(x)))$, by Proposition \ref{dualcor}, the function $\mathfrak{f}=f(h_1,h_2,h_3,h_4)$ is bent and it is given by
 \begin{eqnarray}\label{FF}
 \mathfrak{f}(x,y)=x\cdot \pi(y)\oplus [x\cdot(\pi\oplus \varphi)(y)]\lambda(y)\oplus \eta(y).
 \end{eqnarray}
\end{ex}
\begin{rem}
When $\lambda=0$ (i.e., $g_1=g_2$), then $\mathfrak{f}=h_1\in \mathcal{MM}$. In general, $\mathfrak{f}$  belongs to the set of functions which can be derived from \cite[Theorem 4]{Sihem}. More precisely, if we define bent functions $f_1,f_2,f_3$ as $f_1=h_1$, $f_2=h_2\oplus h_4$, $f_3=h_3$, then in Example \ref{ex:sihem}  we obtain a bent function $\mathfrak{f}$ (relation (\ref{FF})) given as $\mathfrak{f}=f_1 f_2\oplus f_1 f_3\oplus f_2 f_3$, for which the condition $f^*_1\oplus f^*_2\oplus f^*_3=(f_1\oplus f_2\oplus f_3)^*$ can be easily verified using the condition   $\lambda(\pi^{-1}(y))=\lambda(\varphi^{-1}(y))$.
However, when $g_1 \neq g_2$ then  (\ref{FF}) seems not be equivalent  to $f_1f_2\oplus f_1f_3 \oplus f_2f_3$ because our quadratic form $f(x_1,x_2,x_3,x_4)=x_3 (x_2 \oplus x_4) \oplus x_1 (x_2 \oplus x_3 \oplus x_4)=x_1x_2\oplus x_1x_3 \oplus x_2x_3 \oplus (x_3x_4 \oplus x_1x_4)$ involves two additional terms $x_3x_4$ and $x_1x_4$.
\end{rem}

\subsection{Constructions using an indicator set as the form $f$}\label{sec:dual2}

In this section, we provide secondary constructions of bent and plateaued functions (without  increasing the  variable space) by using an indicator set as the form $f$ which essentially simplifies the task of finding suitable initial functions.

We consider the form $f(x_1,\ldots,x_k,x_{k+1})=\phi_{U}(x_1,\ldots,x_k)\oplus x_{k+1}$ defined on $\mathbb{F}^{k+1}_2,$ where $\phi_U$ is the indicator function of an arbitrary subspace $U\subseteq \mathbb{F}^k_2.$ By \cite[Section 3.3.2]{CarletSupp}, it follows that
\[W_{\phi_U}(v)= \left \{ \begin{array}{ll}
2^k-2 \;\#U, & \textnormal{if } v=\textbf{0}_k; \\
-2\;\#U, & \textnormal{if }  v\in U^{\perp}\setminus \{\textbf{0}_k\}; \\
0, & \textnormal{otherwise.}
\end{array} \right.
\]
%
Let us now consider a function $\mathfrak{f}:\mathbb{F}^n_2\rightarrow \mathbb{F}_2$ given by
\begin{eqnarray}\label{indF}
\mathfrak{f}(x)=f(h_1(x),\ldots,h_k(x),a(x))=\phi_U(h_1(x),\ldots,h_k(x))\oplus a(x),\;\; x\in \mathbb{F}^n_2,
\end{eqnarray}
where $a,h_i:\mathbb{F}^{n}_2\rightarrow \mathbb{F}_2$. 
Using the above result it can be easily verified that (for any $v\in \mathbb{F}^n_2$)
\begin{eqnarray}\label{indW}
W_\mathfrak{f}(v)=W_a(v)-2^{1-k}\;\#U\sum_{\omega\in U^{\perp}}W_{a\oplus \omega\cdot (h_1,\ldots,h_k)}(v).
\end{eqnarray}
Notice that compared to (\ref{mainF2}), due to the exact specification of the Walsh spectrum of $\phi_U$, there is no presence of the dual $f^*$ in the computation of $W_\mathfrak{f}(v)$.
\begin{theo}\label{Eth1}
Let $\mathfrak{f}:\mathbb{F}^n_2\rightarrow \mathbb{F}_2$ ($n$ even) be given as $\mathfrak{f}=a\oplus \phi_U(h_1,\ldots,h_k)$, where $a,h_1,\ldots,h_k:\mathbb{F}^n_2\rightarrow \mathbb{F}_2$ and
$\dim(U)=\tau$ ($0\leq \tau\leq k-2$). Assume that $a\oplus \omega\cdot(h_1,\ldots,h_k)$ are bent functions on $\mathbb{F}^n_2$, for all $\omega\in U^{\perp}$. If it holds that $2^{k-\tau}$ divides $\sum_{\omega\in U^{\perp}}(-1)^{(a\oplus \omega\cdot(h_1,\ldots,h_k))^*(v)}$ for all $v\in \mathbb{F}^n_2$,
 then $\mathfrak{f}$ is bent.
\end{theo}
\proof 
Since $\sum_{\omega\in U^{\perp}}(-1)^{(a\oplus \omega\cdot(h_1,\ldots,h_n))^*(v)}=2^{k-\tau}\cdot p$, ($v\in \mathbb{F}^n_2$, $p\in \mathbb{Z}$), we have that
$W_\mathfrak{f}(v)=2^{\frac{n}{2}}((-1)^{a^*(v)}-2p)\equiv 2^{\frac{n}{2}}\;(\text{mod}\;2^{\frac{n}{2}+1})$, which by \cite[Lemma 1]{CarletGPS} implies that $\mathfrak{f}$ is bent.
\qed

The main result of this section is a generic construction method based on the result in Theorem \ref{Eth1}.
\begin{theo}[Generic method A]\label{exEth1}
Let $n$ be even and
 $f_i:\mathbb{F}^n_2\rightarrow\mathbb{F}_2$, for $i \in [1,4]$, be four bent functions such that $f^*_1\oplus f^*_2\oplus f^*_3\oplus f^*_4=0$, where $f_4=f_1\oplus f_2\oplus f_3$.
Taking  $\ell(x)=m\cdot x$ to be any linear function on $\mathbb{F}^n_2$, the function
\begin{equation}\label{eq:sihemext}
\mathfrak{f}(x)=f_1(x)\oplus (\ell\oplus 1)(x)(f_1\oplus f_2\oplus 1)(x)(f_1\oplus f_3\oplus 1)(x),
\end{equation}
on $\F_2^n$ is a bent function.

\end{theo}
\begin{proof}
The proof follows by considering the case $k=3$ and setting  $U=\{\textbf{0}_k\}$ in Theorem \ref{exth-1}, so that  $\dim(U)=\tau=0$ and  $U^{\perp}=\mathbb{F}^3_2$. Let us define  $a,h_1,h_2,h_3:\mathbb{F}^n_2\rightarrow\mathbb{F}_2$ as 
$$a=f_1,\; h_1=\ell,\; h_2=f_1\oplus f_2,\; h_3=f_1\oplus f_3.$$
Then $g_{\omega}=a\oplus \omega\cdot (h_1,h_2,h_3)$ are clearly bent, for all $\omega\in \mathbb{F}^3_2$. In addition, it is not difficult to see that the condition $f^*_1\oplus f^*_2\oplus f^*_3\oplus f^*_4=0$ implies that for some $r \in [0,3]$ (which depends on input variable of $f_i$) it holds that $((-1)^{f^*_1},\ldots,(-1)^{f^*_4})=\pm H_4^{(r)}$, and thus
\begin{eqnarray}\label{heq}
((-1)^{f^*_1},\ldots,(-1)^{f^*_4})=\pm H_4^{(r)}\; \Leftrightarrow\; (-1)^{f^*_1}+\ldots+(-1)^{f^*_4}\in \{0,\pm 4\}.
\end{eqnarray}
For a lexicographically ordered space $\#U^{\perp}=\mathbb{F}^3_2$, by Lemma \ref{l1}-$(i)$, it holds that $\omega_{i+4}=\omega_4\oplus \omega_i$ for all $i\in[0,3]$ $(\omega_i\in \mathbb{F}^3_2)$.  Hence, substituting  $h_1=\ell$, we have that $g_{\omega_{i+4}}=g_{\omega_{i}}\oplus \ell$ and thus $g_{\omega_{i+4}}^*(v)=g_{\omega_{i}}^*(v\oplus m)$  holds for $i\in[0,3]$ and all $v\in \mathbb{F}^n_2$. Consequently, since $(g_{\omega_0},\ldots,g_{\omega_3})=(f_1,f_3,f_2,f_4)$ we have that
 $$\sum_{\omega\in \mathbb{F}^3_2}(-1)^{g_{\omega}^*(v)}=\sum^3_{i=0}(-1)^{g_{\omega_i}^*(v)}+\sum^3_{i=0}(-1)^{g_{\omega_i}^*(v\oplus m)}\stackrel{\textnormal{(\ref{heq})}}{\in}\{0,\pm 2^3\},$$ and thus $2^{k-\tau}=2^3\mid \sum_{\omega\in \mathbb{F}^3_2}(-1)^{g_{\omega}^*(v)}.$

Now, by \cite[Section IV]{Distance},  $\phi_U(x)$ can be represented as $\phi_U(x)=\prod^{k-\tau}_{i=1}(\lambda_i\cdot  x\oplus 1)$, where $\{\lambda_1,\ldots,\lambda_{k-\tau}\}$ is any basis of $U^{\perp}.$ Considering the canonical basis of $U^{\perp}=\mathbb{F}^3_2$, we have that $\phi_U(x)=\prod^{3}_{i=1}(x_i\oplus 1)$. Theorem \ref{Eth1} implies that $\mathfrak{f}=a\oplus \phi_{U}(h_1,h_2,h_3)=a\oplus (h_1\oplus 1)(h_2\oplus 1)(h_3\oplus 1)$ is a bent function whose ANF is then given by (\ref{eq:sihemext}). \qed
\end{proof}
\begin{rem}
Setting $m=\textbf{0}_n$ in Theorem \ref{exEth1} implies that $\mathfrak{f}$ belongs to the set of functions defined in \cite[Theorem 4]{Sihem} whose form is given by $g=f_1f_2 \oplus f_1f_3 \oplus f_2f_3$.  In our case, due to the multiplication of quadratic bent terms $f_if_j$  by a linear function $\ell$, see eq. (\ref{eq:sihemext}), the algebraic degree of $\mathfrak{f}$ is not necessarily the same.
Indeed, by taking larger values of $k$ and employing more linear functions, one can increase the degree of $\mathfrak{f}$ (bounded above by $\frac{n}{2}$) and thus derive affine inequivalent functions to $g$.
\end{rem}
%
Using  the concept of disjoint spectra functions,  where  $f,g\in \mathcal{B}_n$ are said to have disjoint spectra  if $W_f(\omega)W_g(\omega)=0$ for every $\omega\in \mathbb{F}^n_2$ \cite{CCA}, we deduce the following result.
\begin{theo}\label{Eth2}
Let $\mathfrak{f}:\mathbb{F}^n_2\rightarrow \mathbb{F}_2$ be given as $\mathfrak{f}=a\oplus \phi_U(h_1,\ldots,h_k)$, where $a,h_1,\ldots,h_k:\mathbb{F}^n_2\rightarrow \mathbb{F}_2$  and $U$ is a linear subspace
 $\dim(U)=\tau$ with $\tau+2=k$ ($k\geq 2$). Assume that $a\oplus \omega\cdot(h_1,\ldots,h_k)$ are pairwise disjoint spectra $z$-plateaued functions on $\mathbb{F}^n_2$ for all $\omega\in U^{\perp}$, where $z=s+2$ and $0\leq s\leq n-3$. 
Then:
 \begin{enumerate}[i)]
 \item If $z>2$, then $\mathfrak{f}$ is $s$-plateaued.
\item If $n$ is even and $z=2$, then $\mathfrak{f}$ is bent.
\end{enumerate}
\end{theo}
\proof $i)$ Since  $a\oplus \omega\cdot (h_1,\ldots,h_k)$ are pairwise disjoint spectra $z$-plateaued functions on $\mathbb{F}^n_2$, then  we have $\cup_{\omega\in U^{\perp} }S_{a\oplus \omega\cdot (h_1,\ldots,h_k)} \subset \mathbb{F}^n_2$ which follows from the fact that $\#S_{a\oplus \omega\cdot (h_1,\ldots,h_k)}=2^{n-z}<2^{n-2}$ ($z>2$) and  $\#U^{\perp}=2^{k-\tau}=2^2$.
Consequently, there exists  $v\in \mathbb{F}^n_2$ such that $W_{a\oplus \omega\cdot (h_1,\ldots,h_k)}(v)=0$ for all $\omega\in U^{\perp}$, in which case  (\ref{indW}) implies that $W_{\mathfrak{f}}(v)=0$ since $v \not \in S_a$ and thus $W_a(v)=0$. 
Using $\tau+2=k$, we have that the WHT of $\mathfrak{f}$ (at any $v\in \mathbb{F}^n_2$) is given as:
\begin{eqnarray*}
W_{\mathfrak{f}}(v)&=&W_a(v)-2^{1-k+\tau}\sum_{\omega\in U^{\perp}}W_{a\oplus \omega\cdot (h_1,\ldots,h_k)}(v)=W_a(v)-\frac{1}{2}\sum_{\omega\in U^{\perp}}W_{a\oplus \omega\cdot (h_1,\ldots,h_k)}(v)\\
&=&\left\{\begin{array}{cc}
0, & v\not\in S_{a\oplus \omega\cdot (h_1,\ldots,h_k)}\;\text{for all}\; \omega\in U^{\perp},\\
\frac{1}{2}W_a(v)=2^{\frac{n+z}{2}-1}=\pm 2^{\frac{n+s}{2}}, & v\in S_a, \\
-\frac{1}{2}W_{a\oplus \omega'\cdot (h_1,\ldots,h_k)}(v)=\pm 2^{\frac{n+s}{2}}, & v\in S_{a\oplus \omega'\cdot (h_1,\ldots,h_k)},\;\;\omega'\neq \textbf{0}_n.
\end{array}\right.
\end{eqnarray*}

$ii)$ In the case when $n$ is even and $z=2$, then $S_{a\oplus \omega\cdot (h_1,\ldots,h_k)}$ (for $\omega\in U^{\perp}$) partition the space $\mathbb{F}^n_2$ and thus $W_{\mathfrak{f}}(v)\neq 0$ for all $v\in \mathbb{F}^n_2$. In fact, $W_{\mathfrak{f}}(v)$ is given as above for  $s=0$.
\qed

Disjoint spectra plateaued functions satisfying the conditions of Theorem \ref{Eth2} can be constructed efficiently using Theorem \ref{theo:plateH}. We briefly illustrate  the case $i)$ in  Theorem \ref{Eth2}.
\begin{ex}\label{exth-1}
Let $f_1,f_2:\mathbb{F}^n_2\rightarrow \mathbb{F}_2$ be any two $z$-plateaued functions, and let $b$ be any non-zero vector such that the affine subspaces $S_{f_1},S_{f_2},m\oplus S_{f_1}$ and $m\oplus S_{f_2}$ are pairwise disjoint (e.g. take $m\in (S_{f_1}\cup S_{f_2})^{\perp}\subset \mathbb{F}^n_2$). Defining the coordinate functions $a,h_1,h_2:\mathbb{F}^n_2\rightarrow \mathbb{F}_2$ as $$a=f_1,\; h_1=f_1\oplus f_2, \; h_2(x)=m\cdot x, \; x\in \mathbb{F}^n_2,$$ we have that all functions $a\oplus\omega\cdot (h_1,h_2)$ are $z$-plateaued, and considering $U=\{\textbf{0}_2\}$ ($k=2$, $\tau=0$, $U^{\perp}=\mathbb{F}^2_2$), by Theorem \ref{Eth2}-$(i)$ we have that $\mathfrak{f}(x)=f_1(x)\oplus (f_1\oplus f_2)(x)(m\cdot x\oplus 1)$ is an $s$-plateaued function.
\end{ex}
We conclude this section by providing an efficient  generic method that utilizes plateaued functions of different amplitudes (these can be designed using Theorem \ref{theo:plateH}) for specifying new plateaued or bent functions on the same variable space. For this purpose we use the direct sum of linear subspaces $S_{d_i}$ of $\FB^n$, denoted by $S_{d_1} \oplus \cdots \oplus S_{d_r}$.
\begin{prop}[Generic method B]\label{direct}
Let $n$ be even and $d_i:\mathbb{F}^n_2\rightarrow \mathbb{F}_2$ ($i=1,\ldots,r$) be $s_i$-plateaued functions $(2\leq s_i\leq n-2)$ such that $V=S_{d_1} \oplus \cdots \oplus S_{d_r}$,
where each $S_{d_i}$ is a   subspace of $\FB^n$ of dimension $n-s_i$. Assume that  $nr-(s_1+\ldots+s_r)+t=n$, for some $t\geq 0$. Then the function $\mathfrak{f}=d_1\oplus\ldots\oplus d_r$ is $t$-plateaued on $\mathbb{F}^n_2.$ In particular, if $t=0$ then $\mathfrak{f}$ is bent.
\end{prop}
\proof Since the subspace $V$ is a direct sum of Walsh supports $S_{d_i}$ ($s_1+\ldots+s_r=nr+t-n$), then by (\ref{WHT}) for any $u\in\mathbb{F}^n_2$ we have that
$$W_\mathfrak{f}(u)=2^{-nr+\frac{nr+(s_1+\ldots+s_r)}{2}}\sum_{(\omega_1,\ldots,\omega_r)\in S_{d_1}\times\ldots\times S_{d_r}}(-1)^{d^*_1(\omega_1)\oplus\ldots\oplus d^*_r(\omega_r)}\sum_{x\in \mathbb{F}^n_2}(-1)^{(u\oplus \omega_1\oplus\ldots \oplus \omega_r)\cdot x}$$
is equal to $2^{\frac{n+t}{2}}(-1)^{d^*_1(\omega'_1)\oplus\ldots\oplus d^*_r(\omega'_r)}$ if $u\in V$ (since $u$ has a unique representation $u=\omega'_1\oplus\ldots\oplus\omega'_r$ in $V$), or it is equal to $0$ if $u\not\in V$. If $t=0$, then any $u \in V$ and consequently $\mathfrak{f}$ is bent since $W_\mathfrak{f}(u)=\pm 2^{\frac{n}{2}}$. \qed

\section{Conclusions}\label{sec:OP}

%
The compositional representation of Boolean functions  appears to be an efficient tool for deriving secondary constructions of bent/plateaued functions. We derive several explicit design methods some of which are fairly simple and do not involve difficult conditions imposed on the initial functions. The question whether these methods potentially generate bent functions outside the known primary classes (for suitably selected initial functions) remains open.




\begin{thebibliography}{10}




\bibitem{CarletAPN}
{\sc C. Carlet.}
\newblock Boolean and vectorial plateaued functions and APN functions.
\newblock {\em IEEE Transactions on Information Theory}, vol. 61, no. 11, pp. 6272--6289, 2015.

\bibitem{CarletBoolean}
{\sc C. Carlet.}
\newblock Vectorial Boolean functions for cryptography.
\newblock {\em In Y. Crama $\&$ P. Hammer (Eds.), Boolean Models and Methods in Mathematics, Computer Science, and Engineering (Encyclopedia of Mathematics and its Applications), Cambridge University Press}, pp. 398 -- 469, 2010.


\bibitem{CarletGPS}
{\sc C. Carlet.}
\newblock Generalized partial spreads.
\newblock {\em  IEEE Transactions on Information Theory}, vol. 41, no. 5, pp. 1482--1487, 1995.


\bibitem{CarletRes}
{\sc C. Carlet.}
\newblock On bent and highly nonlinear balanced/resilient functions and their algebraic immunities.
\newblock {\em 16th International Symposium, AAECC-16 - Applied Algebra, Algebraic Algorithms and Error-Correcting Code}, LNCS vol. 3857, pp. 1--28 , 2006.

\bibitem{CarletRESB}
{\sc C. Carlet.}
\newblock  On the secondary constructions of resilient and bent functions.
\newblock {\em  Proceedings of the Workshop on Coding, Cryptography and Combinatorics 2003, published by Birkhäuser Verlag}, PCS vol. 23, pp. 3--28, 2004.

\bibitem{CarletOP}
{\sc C. Carlet.}
\newblock Open problems on binary Bent functions.
\newblock {\em  Open Problems in Mathematics and Computational Science - Springer International Publishing}, pp. 203-- 242, 2014.


\bibitem{CarletTNC}
{\sc C. Carlet.}
\newblock Two new classes of bent functions.
\newblock {\em  Advances in Cryptology — EUROCRYPT'93}, LNCS vol. 765, pp. 77--101, 1993.


\bibitem{Decom}
{\sc A. Canteaut, P. Charpin.}
\newblock Decomposing bent functions.
\newblock {\em IEEE Transactions on Information Theory}, vol. 49, no. 8, 2003.

\bibitem{CarletN}
{\sc C. Carlet, H. Dobbertin, G. Leander.}
\newblock Normal extensions of bent functions.
\newblock {\em  IEEE Transactions on Information Theory}, vol. 50, no. 11, 2004.

\bibitem{CarletRSF}
{\sc C. Carlet, G. Gao, W. Liu.}
\newblock A secondary construction and a transformation on rotation symmetric functions, and their action on bent and semi-bent functions.
\newblock {\em  Journal of Combinatorial Theory, Series A}, vol. 127, pp. 161--175, 2014.

\bibitem{CarletMess}
{\sc C. Carlet, S. Mesnager.}
\newblock On Dillon's class $H$ of bent functions, Niho bent functions and o-polynomials.
\newblock {\em  Journal of Combinatorial Theory, Series A}, vol. 118, no. 8, pp. 2392--2410, 2011.

\bibitem{CarletSupp}
{\sc C. Carlet, S. Mesnager.}
\newblock On the supports of the Walsh transforms of Boolean functions.
\newblock {\em  Boolean Functions: Cryptography and Applications, BFCA'05}, pp. 65 -- 82, 2005. Available at:  https://eprint.iacr.org/2004/256.pdf

\bibitem{CarletQ}
{\sc C. Carlet, E. Prouff.}
\newblock On plateaued functions and their constructions.
\newblock {\em  International Workshop on Fast Software Encryption, FSE 2003}, LNCS, vol. 2887, pp. 54--73, 2003.



\bibitem{CarletPiece}
{\sc C. Carlet, J. L. Yucas.}
\newblock Piecewise constructions of bent and almost optimal Boolean functions.
\newblock {\em  Designs, Codes and Cryptography}, vol. 37, no. 3, pp. 449--464, 2005.

\bibitem{SecEnf}
{\sc C. Carlet, F. Zhang, Y. Hu.}
\newblock Secondary constructions of bent functions and their enforcement.
\newblock {\em Advances in Mathematics of Communications}, vol. 6, no. 3, pp. 305 -- 314, 2012.

\bibitem{Ayca2}
{\sc A. \c{C}e\c{s}melio\u{g}lu, G. McGuire, W. Meidl.}
\newblock A construction of weakly and non-weakly regular bent functions.
\newblock {\em Journal of Combinatorial Theory, Series A}, vol. 119, no. 2, pp. 420--429, 2012.

\bibitem{Ayca}
{\sc A. \c{C}e\c{s}melio\u{g}lu, W. Meidl, A. Pott.}
\newblock There are infinitely many bent functions for which the dual is not bent.
\newblock {\em IEEE Transactions on Information Theory}, vol. 62, no. 9, pp. 5204--5208, 2016.








\bibitem{Froben}
{\sc N.~Cepak, E.~Pasalic, A.~ Muratovi\'{c}-Ribi\'{c}.} Frobenius linear translators giving rise to new infinite classes of permutations and bent functions.
\newblock{Available at https://arxiv.org/abs/1801.08460, 2018}

\bibitem{Dillon}
{\sc J. F. Dillon.}
\newblock Elementary Hadamard difference sets. Ph.D. dissertation.
\newblock {\em  University of Maryland, College Park, Md, USA}, 1974.

\bibitem{Dillon2}
{\sc J. F.  Dillon.}
\newblock A survey of bent functions.
\newblock {\em NSA Technical Journal Special Issue}, pp. 191-215, 1972.

\bibitem{Do95}
{\sc H. Dobbertin.}
\newblock  Construction of bent functions and balanced Boolean functions with high nonlinearity.
\newblock {\em Fast Software Encryption '94, Springer-Verlag}, LNCS vol. 1008, pp. 61--74, 1995.

\bibitem{Dobbertin}
{\sc H. Dobbertin, G. Leander, A. Canteaut, C. Carlet, P. Felke, P. Gaborit.}
\newblock  Construction of bent functions via Niho power functions.
\newblock {\em Journal of Combinatorial Theory, Series A}, vol. 113, no. 5, pp. 779-798, 2006.






\bibitem{Xiang}
{\sc X.-D. Hou, P. Langevin.}
\newblock Results on bent functions.
\newblock {\em Journal of Combinatorial Theory, Series A}, vol. 80, no. 2, pp. 232--246, 1997.



\bibitem{Jong}
{\sc J. Y. Hyun, J. Lee, Y. Lee.}
\newblock Explicit criteria for construction of plateaued functions.
\newblock {\em IEEE Transactions on Information Theory }, vol. 62, no. 12, 2016.




\bibitem{Distance}
{\sc  N. Kolomeec, A. Pavlov.}
\newblock  Bent functions on the minimal distance.
\newblock {\em IEEE Region 8 SIBIRCON, Irkutsk Listvyanka, Russia}, July 11 -- 15, 2010.




\bibitem{MMclass}
{\sc R.L. McFarland.}
\newblock A family of noncyclic difference sets.
\newblock {\em Journal of Combinatorial Theory, Series A}, vol. 15, no. 1, pp. 1--10, 1973.

\bibitem{Wilfried}
{\sc W. Meidl.}
\newblock Generalized Rothaus construction and non-weakly regular bent functions.
\newblock {\em Journal of Combinatorial Theory, Series A}, vol. 141, pp. 78--89, 2016.

\bibitem{SihemBook}
{\sc S. Mesnager.}
\newblock Bent Functions - fundamentals and results.
\newblock {\em Theoretical Computer Science - Springer International Publishing}, 2016.

\bibitem{Sihem2}
{\sc S. Mesnager.}
\newblock Further constructions of infinite families of bent functions from new permutations and their duals.
\newblock {\em Cryptography and Communications}, vol. 8, no. 2, pp. 229--246, 2016.

\bibitem{Sihem}
{\sc S. Mesnager.}
\newblock Several new infinite families of bent functions and their duals.
\newblock {\em IEEE Transactions on Information Theory}, vol. 60, no. 7, pp. 4397--4407, 2014.

\bibitem{SihemN}
{\sc S.~Mesnager, P.~Ongan, F.~\"{O}zbudak.}
\newblock New bent functions from permutations and linear translators.
\newblock {\em  Codes, Cryptology and Information Security}, LNCS vol. 10194, pp. 282--297, 2017.

\bibitem{SihemPlate}
{\sc S. Mesnager, F. {\"O}zbudak, A. Sinak.}
\newblock Results on characterizations of plateaued functions in arbitrary characteristic.
\newblock {\em Cryptography and Information Security in the Balkans}, LNCS vol. 9540, pp. 17--30, 2016.








\bibitem{Rot}
{\sc O. S. Rothaus.}
\newblock On "bent" functions.
\newblock {\em Journal of Combinatorial Theory, Series A}, vol. 20, no. 3, pp. 300--305, 1976.

\bibitem{CCA}
{\sc P. Sarkar, S. Maitra.}
\newblock Cross-correlation analysis of cryptographically useful Boolean functions and S-boxes.
\newblock {\em Theory of Computing Systems}, vol. 35, no. 1, pp. 39--57, 2002.

\bibitem{SeberryX}
{\sc J. Seberry, X-M. Zhang.}
\newblock Highly nonlinear $0-1$ balanced Boolean functions satisfying strict avalanche criterion.
\newblock {\em Advances in Cryptography - Auscrypt'92, Berlin-Springer}, LNCS vol. 718, pp. 145--755, 1993.









\bibitem{Wang}
{\sc W. Weiqiong, X. Guozhen.}
\newblock Decomposition and construction of plateaued functions.
\newblock {\em Chinese Journal of Electronics}, vol. 18, no. 4, 2009.




\bibitem{Feng2}
{\sc F. Zhang, C. Carlet, Y. Hu, T.-J. Cao.}
\newblock Secondary constructions of highly nonlinear Boolean functions and disjoint spectra plateaued functions.
\newblock {\em  Information Sciences}, vol. 283, pp. 94--106, 2014.

\bibitem{Feng}
{\sc F. Zhang, C. Carlet, Y. Hu, W. Zhang.}
\newblock New secondary constructions of bent functions.
\newblock {\em  Applicable Algebra in Engineering, Communication and Computing}, vol. 27, no. 5, pp. 413--434, 2016.


\bibitem{RothEnes2016}
{\sc F. Zhang, E. Pasalic, Y. Wei, N. Cepak.}
\newblock Constructing bent functions outside the Maiorana-McFarland class using a general form of Rothaus.
\newblock {\em  IEEE Transactions on Information Theory}, vol. 63, no. 8, pp. 5336 -- 5349, 2017.

\bibitem{OutsideMM}
{\sc F. Zhang, E. Pasalic, N. Cepak, Y. Wei}
\newblock Bent functions in $\mathcal{C}$ and $\mathcal{D}$ outside the completed Maiorana-McFarland class.
\newblock {\em Codes, Cryptology and Information Security}, C2SI, LNCS 10194, Springer-Verlag, pp. 298--313 (2017).

\bibitem{Feng4}
{\sc F. Zhang, Y. Wei, E. Pasalic.}
\newblock Constructions of bent - negabent functions and their relation to the completed Maiorana - McFarland class.
\newblock {\em  IEEE Transactions on Information Theory}, vol. 61,  no. 3, pp. 1496--1506, 2015.


\bibitem{Zheng}
{\sc Y. Zheng, X. M. Zhang.}
\newblock On plateaued functions.
\newblock {\em IEEE Transactions on Information Theory}, vol. 47, no. 3, pp. 1215--1223, 2001.

\bibitem{Zheng2}
{\sc Y. Zheng, X. M. Zhang.}
\newblock Relationships between bent functions and complementary plateaued functions.
\newblock {\em Information Security and Cryptology - ICISC 99, Springer, Berlin}, LNCS vol. 1787, pp. 60--75, 2000.



\end{thebibliography}
\end{document}